\documentclass[manuscript]{emulateapj}
\usepackage{amsmath}
\usepackage[usenames,dvipsnames]{color}
\usepackage{graphicx}
\usepackage{natbib}
\usepackage{times}
\usepackage{enumitem}

\newcommand{\cena}{NGC\,5128}

\newcommand{\hdyn}{DSC}
\newcommand{\ldyn}{DGTO}

\slugcomment{Submitted to The Astrophysical Journal}

\shorttitle{The Dark Side of \cena's Globular Clusters}
\shortauthors{M. A. Taylor et al.}

\begin{document}

\title{Observational Evidence for a Dark Side to NGC 5128's Globular Cluster System\footnotemark[$\dagger$]} 
\footnotetext[$\dagger$]{Based on observations collected under program 081.D-0651 (PI: Matias Gomez) with FLAMES at the Very Large Telescope of the Paranal Observatory in Chile, operated by the European Southern Observatory (ESO).}

\author{Matthew A. Taylor$^{1,2,\star}$, Thomas H.~Puzia$^{1}$, Matias Gomez$^{3}$, Kristin A. Woodley$^{4}$}
\affil{
$^{1}$Institute of Astrophysics, Pontificia Universidad Cat\'olica de Chile, Av.~Vicu\~na Mackenna 4860, 7820436 Macul, Santiago, Chile\\
$^{2}$European Southern Observatory, Alonso de Cordova 3107, Vitacura, Santiago, Chile\\
$^{3}$Departamento de Ciencias F\'isicas, Universidad Andres Bello, Rep\'ublica 220, Santiago, Chile\\
$^{4}$University of California, Santa Cruz, University of California Observatories, 1156 High Street, Santa Cruz, CA 95064, USA}
\footnotetext[$\star$]{ESO Graduate Student Fellow, {\it mtaylor@astro.puc.cl}}

\begin{abstract}
We present a study of the dynamical properties of 125 compact stellar systems (CSSs) in the nearby giant elliptical galaxy \cena, using high-resolution spectra ($R\!\approx\!26\,000$) obtained with VLT/FLAMES. Our results provide evidence for a new type of star cluster, based on the CSS dynamical mass scaling relations.~All radial velocity ($v_r$) and line-of-sight velocity dispersion ($\sigma_{\rm los}$) measurements are performed with the penalized pixel fitting ({\it ppxf}) technique, which provided $\sigma_{\rm ppxf}$ estimates for 115 targets.~The $\sigma_{\rm ppxf}$ estimates are corrected to the 2D projected half-light radii, $\sigma_{1/2}$, as well as the cluster cores, $\sigma_0$, accounting for observational/aperture effects and are combined with structural parameters, from high spatial resolution imaging, in order to derive total dynamical masses (${\cal{M}}_{\rm dyn}$) for 112 members of \cena's star cluster system.~In total, 89 CSSs have dynamical masses measured for the first time along with the corresponding dynamical mass-to-light ratios ($\Upsilon_V^{\rm dyn}$).~We find two distinct sequences in the $\Upsilon_V^{\rm dyn}$-${\cal{M}}_{\rm dyn}$ plane, which are well approximated by power laws of the forms $\Upsilon_V^{\rm dyn}\!\propto\!{\cal{M}}_{\rm dyn}^{0.33\pm0.04}$ and $\Upsilon_V^{\rm dyn}\!\propto\!{\cal{M}}_{\rm dyn}^{0.79\pm0.04}$.~The shallower sequence corresponds to the very bright tail of the globular cluster luminosity function (GCLF), while the steeper relation appears to be populated by a distinct group of objects which require significant dark gravitating components such as central massive black holes and/or exotically concentrated dark matter distributions. This result would suggest that the formation and evolution of these CSSs are markedly different from the ``classical" globular clusters in \cena\ and the Local Group, despite the fact that these clusters have luminosities similar to the GCLF turn-over magnitude.~We include a thorough discussion of myriad factors potentially influencing our measurements.
\end{abstract}


\section{Introduction}
\label{sec:intro}
Globular clusters (GCs) are among the oldest stellar systems in the Universe \citep{kra03}.~They have witnessed the earliest stages of star formation and were also present during later epochs of structure formation.~Apart from resolved stellar population studies of galaxies, which are restricted primarily to the Local Group, extragalactic globular cluster systems (GCSs) provide one of the best probes to investigate the formation and assembly histories of galaxies \citep{har91, ash98, ash08, pen08, geo10}.~Various avenues of study can be employed to this effect, including the analysis of GCS kinematics, their metallicity distribution functions, chemical enrichment histories, age spreads, and/or combinations thereof.

At a distance of $3.8\!\pm\!0.1$\,Mpc, \citep[][]{har10}, corresponding to an angular scale of 18.5\,pc\,arcsec$^{-1}$, \cena\ (a.k.a.~Centarus~A) is the nearest giant elliptical (gE) galaxy to the Milky Way (MW), yet it is still too far for large scale star-by-star investigations to be technologically feasible.~Fortunately, much has been learned about this galaxy from its rich GC system. 564 of its GCs have radial velocity confirmations, and others have been confirmed, for example, via resolution into individual stars \citep{van81, hes84, hes86, har92, pen04, woo05, rej07, bea08, woo10a}.~This GC sample, thus, rivals the entire population of GCs harbored by the Local Group, despite there being $\sim\!600\!-\!1400$ GCs still left to find/confirm in the halo regions of \cena\ \citep{har84, har02b, har10, har12}.~Notwithstanding this incompleteness, previous studies have already shed much light on the photometric, chemical, and kinematical properties, as well as the past and recent formation history of this massive nearby neighbour.

GCs are well known to inhabit a narrow range of space defined by structural parameters such as half-light, tidal and core radii ($r_h$, $r_t$ and $r_c$, respectively), concentration parameter $c\!=\!\log\left(r_t/r_c\right)$ \citep{kin66}, velocity dispersion ($\sigma$), mass-to-light ratio ($\Upsilon$), etc.~called the ``fundamental plane'' \citep{djo95}.~Studies of Local Group GCs \citep[e.g.][]{fus94, djo97, hol97, mcl00, bar02, bar07} have shown that at the high-mass end of the fundamental plane, peculiar GCs such as $\omega$Cen and G1, the largest GCs in the MW and M31, begin to emerge.~For example, both of these GCs show significant star-to-star [Fe/H] variations and are among the most flattened of Local Group GCs \citep{whi87, nor95, mey01, pan02} and at least in the case of $\omega$Cen, harbor multiple stellar populations with an extended chemical enrichment history \citep{pio08a, pio08b}.

Unlike the Local Group, the sheer size of \cena's GCS generously samples the high-mass tail ($\ga\!10^6$ $M_\odot$) of the globular cluster mass function \citep[GCMF,][]{har84, har02a, mar04, rej07, tay10}.~Due to their intense luminosities, these massive GCs are very accessible observationally, and thus provide excellent probes to study the formation history of \cena.~Many of the most massive \cena\ GCs show a more rapid chemical enrichment history than Local Group GCs \citep{col13}, and exhibit significantly elevated dynamical mass-to-light ratios ($\Upsilon_V^{\rm dyn}$) above dynamical masses, ${\cal{M}}_{\rm dyn}\simeq2\cdot10^6$ $M_\odot$ \citep{tay10}.~This sharp upturn of $\Upsilon_V^{\rm dyn}$ is consistent with a trend found by \cite{has05} and \cite{mie06, mie08a} in other extragalactic GCSs, and requires either non-equilibrium dynamical states, such as rotation or pre-relaxation \citep{varri12, bia13}, younger than expected stellar components \citep[e.g.][]{bed04,pio08a}, exotic top- or bottom-heavy stellar initial mass functions \citep[IMFs; e.g.][]{dab08, dab09, mie08b} or/and a significant contribution by non-baryonic matter or massive central black holes (BHs).~While there is an ongoing debate whether the latter two options are valid for Milky Way GCs \citep[see e.g.][]{con11, iba13, lue11, str12, lan13, sun13, kru13}, there is growing evidence for the presence of $10^5-10^8\,M_\odot$ BHs significantly affecting the dynamics of similarly structured, albeit more massive, ultra-compact dwarf galaxies \citep[UCDs;][]{mie13,set14}.

\cena's GCS has been shown to follow trends similar to other giant galaxies.~In particular, it has a multi-modal distribution in color and metallicity \citep[e.g.][]{har02b, pen04, bea08, woo10b}, corresponding to at least two and possibly three distinct GC generations.~Moreover, the prominent dust-lane and faint shells in the galaxy surface brightness distribution \citep{mal83}, along with a young tidal stream \citep{pen02} provide significant evidence for recent merger activity on a kpc scale, while on smaller scales indications of strong tidal forces are seen in the form of extra-tidal light associated with individual GCs \citep{har02a}.

Recent models support the notion that the bulk of the star formation leading to massive elliptical galaxies is complete by $z\!\approx\!3$ (i.e.~the first few Gyr of cosmic history), while it takes until $z\!<\!0.4$ before $\sim\!80\%$ of the mass is locked up after the accretion of as many as five massive progenitors \citep[e.g.][]{del06,del07,mar14}.~Studies based on {\it Hubble Space Telescope} (HST) data in the mid- to outer-halo regions of \cena\ generally concur with this view, in that the majority ($\sim80\%$) of the stellar population is ancient ($\ga\!11\!-\!12$ Gyr) and formed very rapidly, as evidenced by [$\alpha$/Fe] ratios approaching or exceeding twice solar values \citep[e.g.][]{har99,2har00,2har02,rej11}.~This older population is complemented by a significantly younger component, forming on the order of a few Gyr ago \citep[e.g.][]{sor96,mar00,rej03}.

In this paper we use velocity dispersion estimates based on high-resolution spectra to derive dynamical masses for a large sample of \cena's GCS \citep[see e.g.,][who carried out similar work on ultra-compact dwarfs in the Fornax cluster]{chi11}. We combine the newly derived dynamical information with well-known luminosities from the literature to probe the baryonic makeup and dynamical configurations of the CSSs. The results are then used to classify several distinct CSS/GC populations, which are discussed in the context of likely origins, with potential consequences for GCSs that surround other gE galaxies.

This paper is organized as follows.~\S~\ref{sec:obs} describes the observations made as well as an outline of the data reduction steps taken to produce high-quality spectra.~\S~\ref{sec:analysis} contains information on the analysis that was undertaken on the new spectroscopic observations, as well as structural parameter data from the literature with which our new measurements were combined.~\S~\ref{sec:disc} discusses our results by using sizes, masses and mass-to-light ratios of GCs/CSSs to develop several hypotheses on the origins of the various cluster sub-populations that we find. The main text concludes with \S~\ref{sec:conc}, which summarizes our new measurements and results.~Following the main text, we present in the appendix multiple detailed tests which rule out spurious results due to several possible sources including poor data quality, data analysis biases, fore/background contamination, target confusion, and others.~We adopt the \cena\ distance modulus of $(m\!-\!M)_0\!=\!27.88\pm0.05$ mag, corresponding to a distance of $3.8\pm0.1$ Mpc \citep{har10}, as well as the homogenized GC identification scheme of \cite{woo07} throughout this work.

\section{Observations}
\label{sec:obs}

\begin{figure*}
\centering
\includegraphics[width=17.5cm]{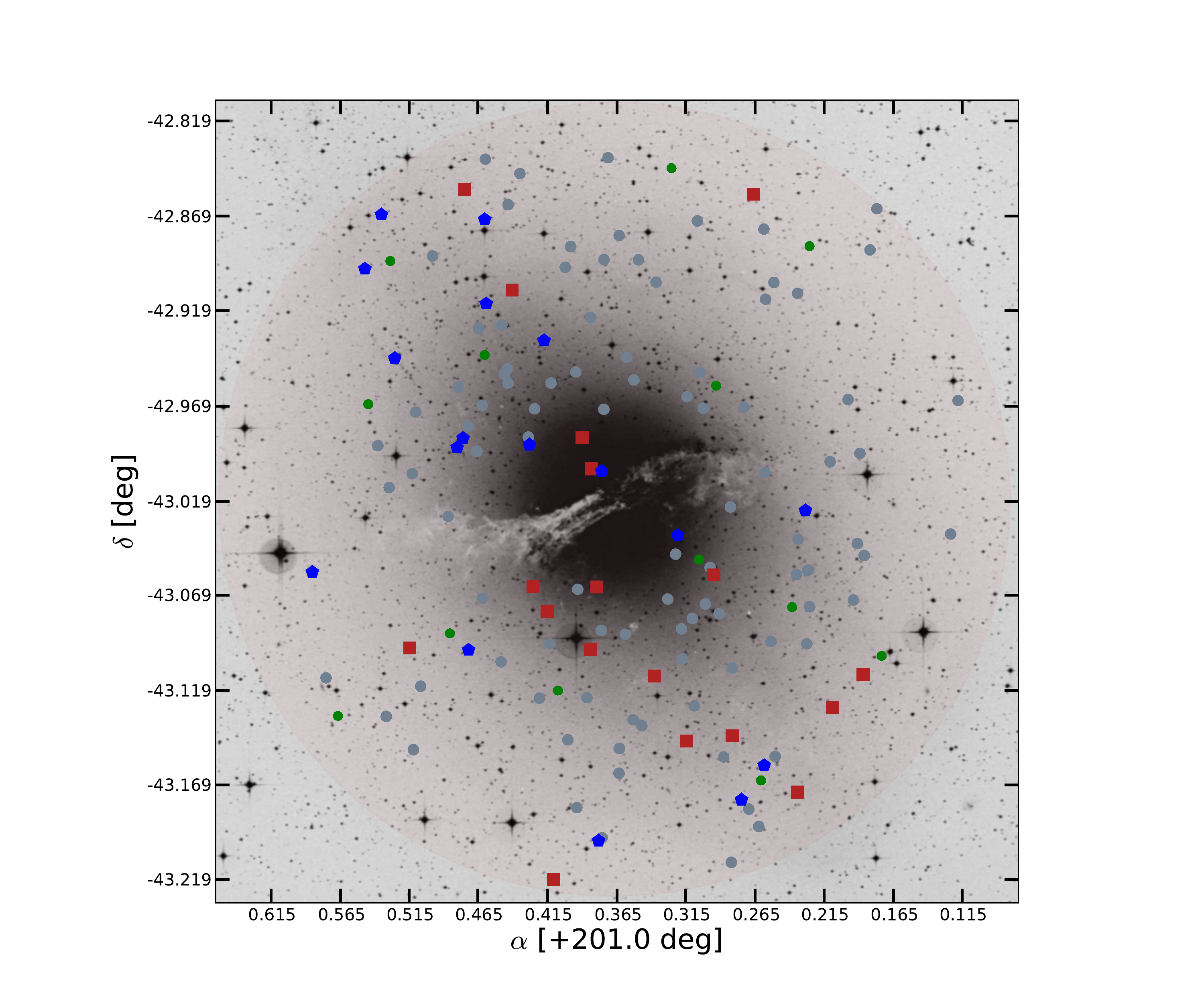}
\caption{Locations of the GIRAFFE fibre placements in relation to the host galaxy, \cena. Grey points show the placement of fibres for ``classical" GCs, while blue pentagons and red squares illustrate the positions of \ldyn\ and \hdyn\ GCs, respectively (see Figure~\ref{fig:mass_mld_size} and corresponding text in Sect.~\ref{sec:md_v_ml} for details on their definition).~Green points mark the positions of sky fibres.~The light red shaded region shows the approximate extent of the 25\arcmin\ FLAMES field-of-view, corresponding to $\sim27.5$ kpc at the distance of \cena.}
\label{fig:n5128_targs}
\end{figure*} 

During five nights in June/July 2008, 123 of the brightest GCs around \cena\ were observed using the Fibre Large Array Multi-Element Spectrograph (FLAMES) instrument at the Very Large Telescope (VLT) on Cerro Paranal, Chile. FLAMES is a multi-object spectrograph mounted at the Nasmyth A focus of UT2 (Kueyen). The instrument features 132 fibres, each with apertures of 1.2\arcsec\ diameter, linked to the intermediate-high resolution ($7\,000\!\leq\!R\!\leq\!30\,000$) GIRAFFE spectrograph, with an additional eight 1.0\arcsec\ fibres connected to the high resolution ($R\!\approx\!47\,000$) UVES spectrograph mounted at the Nasmyth B focus, thus, allowing for simultaneous observations of 139 targets\footnote{In principle 132 GIRAFFE fibres are allocatable, but only 131 are fully covered on the detector; therefore, the sum of GIRAFFE+UVES available fibres is 139.} over a 25\arcmin\ diameter field of view. 

\subsection{Instrumental Setup}
FLAMES often suffers from one or more broken fibres, and these observations were no exception.~The first night of observations, comprising the first three of eleven 2\,400 second long observing blocks (OBs), were conducted with the first of two FLAMES fibre positioner plates, which suffered from two broken fibres, while the last eight OBs used the second plate, with only a single, but distinct, broken fibre.~For this reason, the total number of targets observed was 138, including eight fibres fed to UVES, and 130 to GIRAFFE.~Unfortunately the UVES targets were of too low data quality for useful measurements to be derived, and so we do not include them in the present work. From the GIRAFFE fibre budget, 13 were allocated to recording the sky contribution to the target signals.~These sky fibres were used near the end of the data reduction process to perform an adaptive sky subtraction.~In summary, we obtained spectra of 117 bright GCs using GIRAFFE in the high resolution ($R\!\simeq\!25\,900$) mode, recording a single $\Delta\lambda = 21.3$\,nm echelle order centred at $\lambda\!=\!525.8$\,nm, covering the wavelength range $515.2\la\lambda\la536.5\,{\rm nm}$.

Table~\ref{tbl:observations} summarizes our observations.~The overall exposure times were 26\,400\,s; however, three GCs (GC\,0310, GC\,0316 and GC\,426) inevitably suffered from broken fibres, limiting the total integration times to 19\,200\,s, 7\,200\,s, and 7\,200\,s, respectively.~In the case of GC\,0058 a single exposure had to be omitted due to the unfortunate coincidence of a significant detector defect lying directly in the middle of the Mg$b$ triplet, hence the total integration time for that GC was limited to 24\,000\,s.

The on-sky locations of the GIRAFFE fibres are indica\-ted in Figure~\ref{fig:n5128_targs}, over-plotted on an archival DSS image\footnote{Based on photographic data obtained using The UK Schmidt Telescope. The UK Schmidt Telescope was operated by the Royal Observatory Edinburgh, with funding from the UK Science and Engineering Research Council, until 1988 June, and thereafter by the Anglo-Australian Observatory. Original plate material is copyright \copyright\ the Royal Observatory Edinburgh and the Anglo-Australian Observatory. The plates were processed into the present compressed digital form with their permission. The Digitized Sky Survey was produced at the Space Telescope Science Institute under US Government grant NAG W-2166.} of \cena.~Overall, the observing conditions were good for this observing program:~The images were taken at airmass values ranging between 1.054 and 1.635, under seeing conditions in the range 0.48\arcsec\ to 1.42\arcsec.~For the purpose of correcting our line-of-sight velocity dispersion measurements, based on our final stacked spectra, for aperture and observational effects (see \S~\ref{sec:apcorr}), we use the mean seeing value from all 11 OBs of 0.85\arcsec, since all targets were observed simultaneously under identical seeing conditions.

\begin{deluxetable*}{lcccccc}
\tabletypesize{\scriptsize}
\tablecaption{Star Cluster Observations\label{tbl:observations}}
\tablewidth{0pt}
\tablehead{
\colhead{ID} & \colhead{$\alpha$} & \colhead{$\delta$} & \colhead{$R$} & \colhead{$V_0$} & \colhead{Exp.~Time} & \colhead{S/N} \\
\colhead{} & \colhead{[J2000]} & \colhead{[J2000]} & \colhead{[mag]} & \colhead{[mag]} & \colhead{[s]} & }
\startdata
GC\,0028 & 13 24 28.429 & $-42$ 57 52.96 & 19.65 & 19.80$\pm$0.01 & 26400 & 2.87 \\
GC\,0031 & 13 24 29.700 & $-43$ 02 06.43 & 19.55 & 19.75$\pm$0.01 & 26400 & 1.88 \\
GC\,0048 & 13 24 43.586 & $-42$ 53 07.22 & 19.33 & 19.51$\pm$0.01 & 26400 & 1.79 \\
GC\,0050 & 13 24 44.575 & $-43$ 02 47.26 & 18.90 & 18.74$\pm$0.01 & 26400 & 3.75 \\
GC\,0052 & 13 24 45.330 & $-42$ 59 33.47 & 18.91 & 18.98$\pm$0.01 & 26400 & 5.55 \\
GC\,0053 & 13 24 45.754 & $-43$ 02 24.50 & 19.43 & 19.57$\pm$0.01 & 26400 & 2.22 \\
GC\,0054 & 13 24 46.435 & $-43$ 04 11.60 & 18.64 & 18.84$\pm$0.01 & 26400 & 4.65 \\
GC\,0058 & 13 24 47.369 & $-42$ 57 51.19 & 19.15 & 19.41$\pm$0.01 & 24000 & 2.69 \\
GC\,0064 & 13 24 50.072 & $-43$ 07 36.23 & 20.03 & 20.11$\pm$0.02 & 26400 & 1.49 \\
GC\,0065 & 13 24 50.457 & $-42$ 59 48.98 & 19.68 & 19.21$\pm$0.01 & 26400 & 2.17
\enddata
\tablecomments{Summary of the new observations. Cluster identifications are listed in the first column, followed by the J2000 coordinates, apparent $R$-band magnitudes used for target acquisitions, de-reddened apparent $V$- or $r'$-band magnitudes (see \S\ref{sec:masses}), total integration times, and signal-to-noise ratios (S/N, see Sect.~\ref{sec:datreduc} for a definition). Table~\ref{tbl:observations} is published in its entirety in the electronic edition of the {\it Astrophysical Journal}. A portion is shown here for guidance regarding its form and content.}
\end{deluxetable*}

\subsection{Basic Data Reduction and Cleaning}
\label{sec:datreduc}
The basic data reduction steps (bias subtraction, flat fielding, and wavelength calibration) were carried out by the GIRAFFE pipeline\footnote{\url{http://www.eso.org/sci/software/pipelines}}.~Separate calibration frame sets were used for each of the five nights.~The pipeline recipe {\it masterbias} created the master bias frame from an average of five individual frames and {\it masterflat} produced the master flat from an average of three bias-subtracted flatfields.~The fibre localizations were visually confirmed to be accurate to within 0.5 pixels for each of the 11 OBs, less than the suggested 1 pixel maximum to ensure accuracy.~The recipe {\it giwavecalibration} derived the wavelength calibrations.~For all five calibration sets, it was necessary to edit the slit geometry tables in order to eliminate `jumps' in the final, re-binned, wavelength calibrated arc-lamp spectra -- an extra step that is not uncommon.~Having performed these steps, the wavelength solutions were confirmed to be of high-quality by visually checking that they were smooth, as well as via the radial velocity errors internal to the re-made slit geometry tables which showed values of ${\it RVERR}\simeq0.003$\,km s$^{-1}$. We note that these values are meant to confirm the accuracy of {\it giwavecalibration} and do not reflect our final, measured radial velocity uncertainties (see \S~\ref{sec:analysis}).

Using the final 11 sets of calibration data products, the recipe {\it giscience} provided the final, fully calibrated science frames from which individual 1D spectra were extracted.~Custom {\sc Python} scripts were used to clean the spectra of numerous residual cosmetic defects and to subtract the sky contribution from the spectra.~To clean the spectra of cosmetic defects surviving the basic data reduction steps, the spectra were subjected to a median filtering algorithm and robust $\kappa\sigma$-clipping.~Each of the extracted spectra were visually inspected and the parameters of the median/$\kappa\sigma$ filters were adjusted to remove any significant detector cosmetics, while preserving the finer details of the spectra.~Typically a median filter of gate size of 75 pixels followed by clipping points outside of 4.5$\sigma$ was sufficient to remove defects.

\subsection{Sky Subtraction}
\label{sec:skysub}
To account for the sky contribution to each spectrum, we used the 13 GIRAFFE fibres dedicated to monitoring the sky contamination.~These fibres facilitated uniform sampling across the field of view (see Figure~\ref{fig:n5128_targs}).~For each target, the sky contribution was taken to be the average of the three nearest sky fibres, inversely weighted by distance, thereby ensuring that only the sky nearest to each target was considered.~The final sky spectra were determined individually for each of the 117 targets and 11 OBs before being directly subtracted from each of the 1287 individual target spectra.~Only then were the reduced, cleaned, and sky-subtracted spectra co-added to produce the final data set.

\subsection{Data Quality Assessment}
The signal-to-noise ratios (S/N) for the final spectra were calculated considering the main spectral features used to estimate the line-of-sight velocity dispersions (see \S\ref{sec:ppxf}).~Specifically, these features are the Mg$b$ and Fe\,5270 Lick indices centred at laboratory wavelengths of 5176.375\,\AA\ and 5265.650\,\AA, respectively \citep{bur84,wor94,wor97}.~The S/N listed in Table~\ref{tbl:observations} were calculated as,
\begin{equation}
S/N = \frac{1}{4}\sum_i^4{\frac{s_i}{\sigma_i}}
\end{equation}
where $s_i$ and $\sigma_i$ are the mean and standard deviation of the flux over the continuum regions bracing the Mg$b$ and Fe\,5270 features as defined by $5142.625\,{\rm \AA}\!\leq\lambda_{{\rm Mg}b,{\rm cont}}^{\rm blue}\!\leq\!5161.375\,{\rm \AA}$, $5191.375\,{\rm \AA}\!\leq\lambda_{{\rm Mg}b,{\rm cont}}^{\rm red}\!\leq\!5206.375\,{\rm \AA}$, $5233.150\,{\rm \AA}\!\leq\lambda_{{\rm Fe5270},{\rm cont}}^{\rm blue}\!\leq\!5248.150\,{\rm \AA}$ and $5285.650\,{\rm \AA}\!\leq\lambda_{{\rm Fe5270},{\rm cont}}^{\rm red}\!\leq\!5318.150\,{\rm \AA}$.~Before calculating the S/N, each of the continuum definitions were shifted from the laboratory values to account for known GC radial velocities, $v_r$, or if unknown, they were shifted {\it a posteriori} according to our own $v_r$ measurements (see \S~\ref{sec:ppxf}).

\section{Analysis}
\label{sec:analysis}
\subsection{Penalized Pixel Fitting}
\label{sec:ppxf}
Our line-of-sight velocity dispersion (LOSVD; $\sigma$) measurements were carried out using the penalized pixel fitting ({\it ppxf}) code \citep{cap04}.~This code parametrically recovers the LOSVD of the stars composing a given cluster or galaxy spectrum by expanding the LOSVD profile as a Gauss-Hermite series.~Using reasonable initial guesses for the radial velocity ($v_r$) and $\sigma$, the best fitting $v_{r,{\rm ppxf}}$, $\sigma_{\rm ppxf}$, and Hermite moments $h_3,$ and $h_4$ were recovered by fitting the cluster/galaxy spectrum to a library of template stars which had its spectral resolution adjusted to that of the FLAMES spectra.~The fitting of optimal template spectra along with the kinematics serves to limit the impact of template mismatches.~An important feature of the {\it ppxf} routine is that during an iterative process, a penalty function derived from the integrated square deviation of the line profile from the best fitting Gaussian is used to minimize the variance of the fit.~This feature allows the code to recover the higher order details in high S/N spectra, but biases the solution towards a Gaussian when S/N is low, as is the case for several objects in our sample.~For more details on the {\it ppxf} code, we refer to \cite{cap04}\footnote{{\it ppxf} and the corresponding documentation can be found at: \url{http://www-astro.physics.ox.ac.uk/\~{}mxc/idl/}}.

Where possible, the input estimates for $v_r$ were quoted from \cite{woo10a} or \cite{woo07} which are listed in Table~\ref{tbl:kinematics}.~For GCs with unavailable $v_r$, we used the IRAF\footnote{IRAF is distributed by the National Optical Astronomy Observatory, which is operated by the Association of Universities for Research in Astronomy, Inc., under cooperative agreement with the National Science Foundation.} task {\it rvcorrect} to account for heliocentric velocity corrections, and {\it fxcor} to estimate $v_r$.~These estimates were used as our initial guesses for {\it ppxf} and do not need to be perfectly accurate since the $\sigma_{\rm ppxf}$ measurements have no significant sensitivity to $v_r$, as long as the initial guess is accurate to within a few tens of km s$^{-1}$ \citep[see also][]{tay10}.~The more refined $v_{r,{\rm ppxf}}$ values were then adopted as our final estimates as listed in Table~\ref{tbl:kinematics} and used for our S/N measurements.

\begin{figure}[t]
\centering
\includegraphics[width=8.9cm]{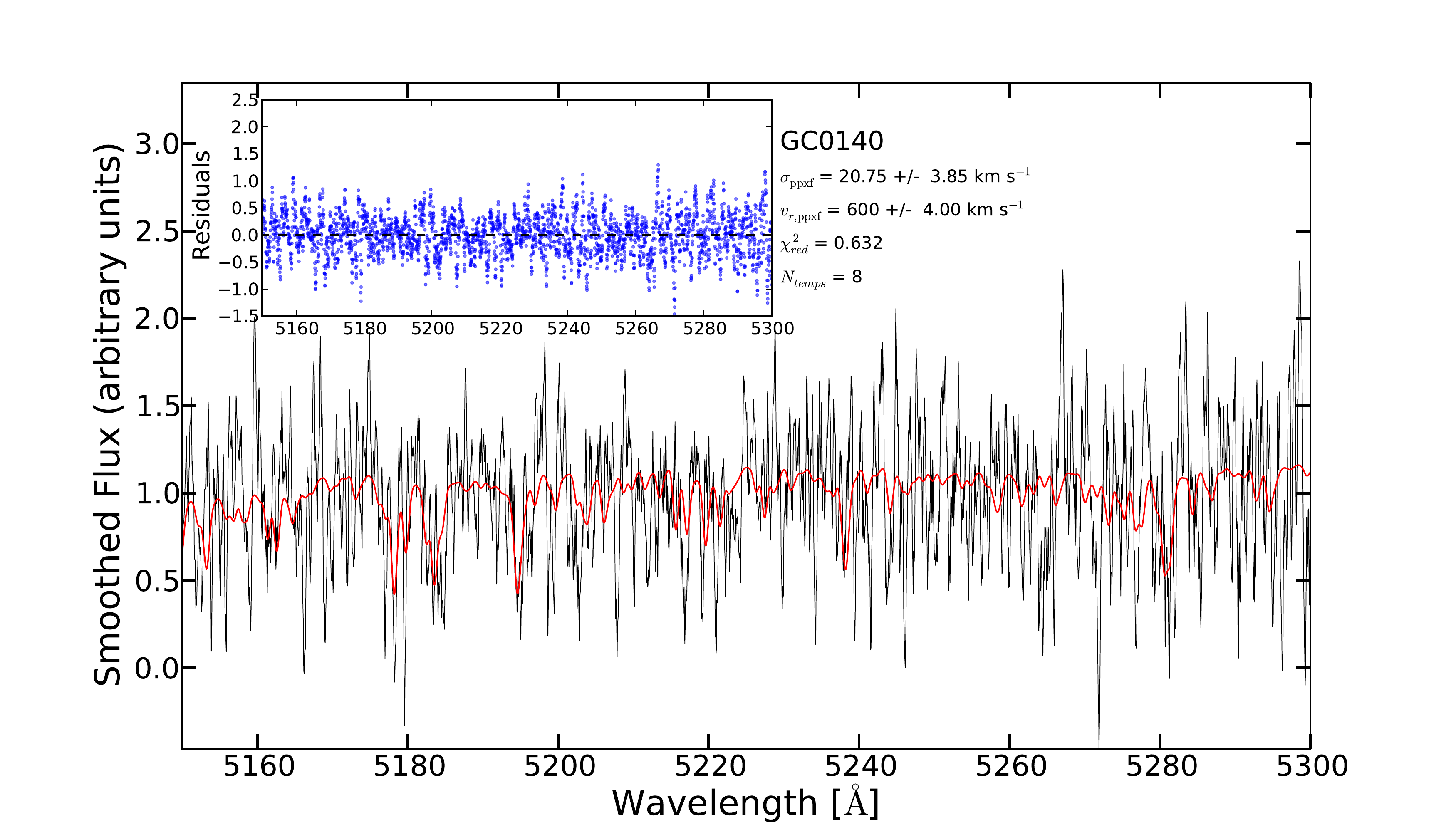}
\includegraphics[width=8.9cm]{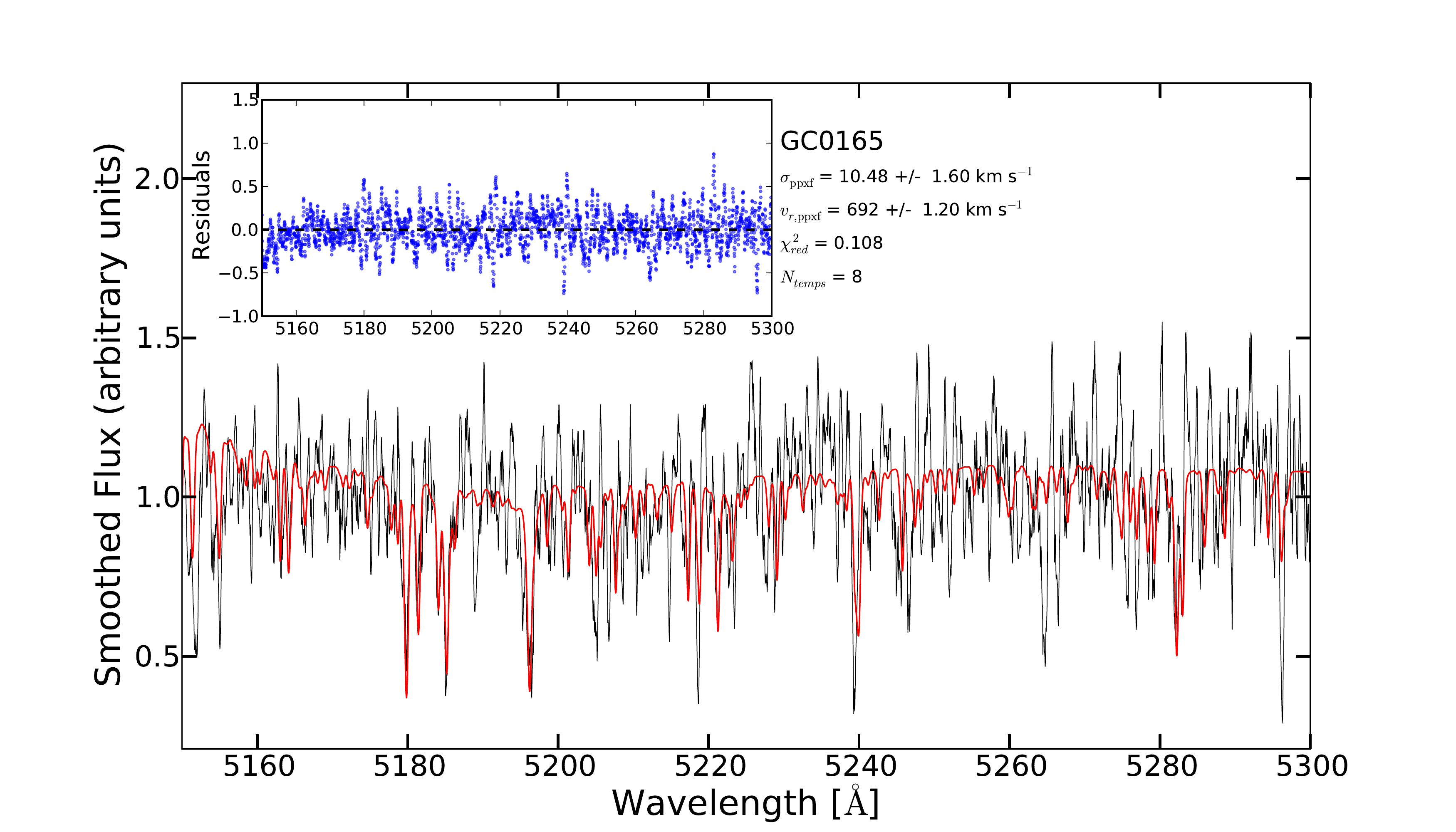}
\includegraphics[width=8.9cm]{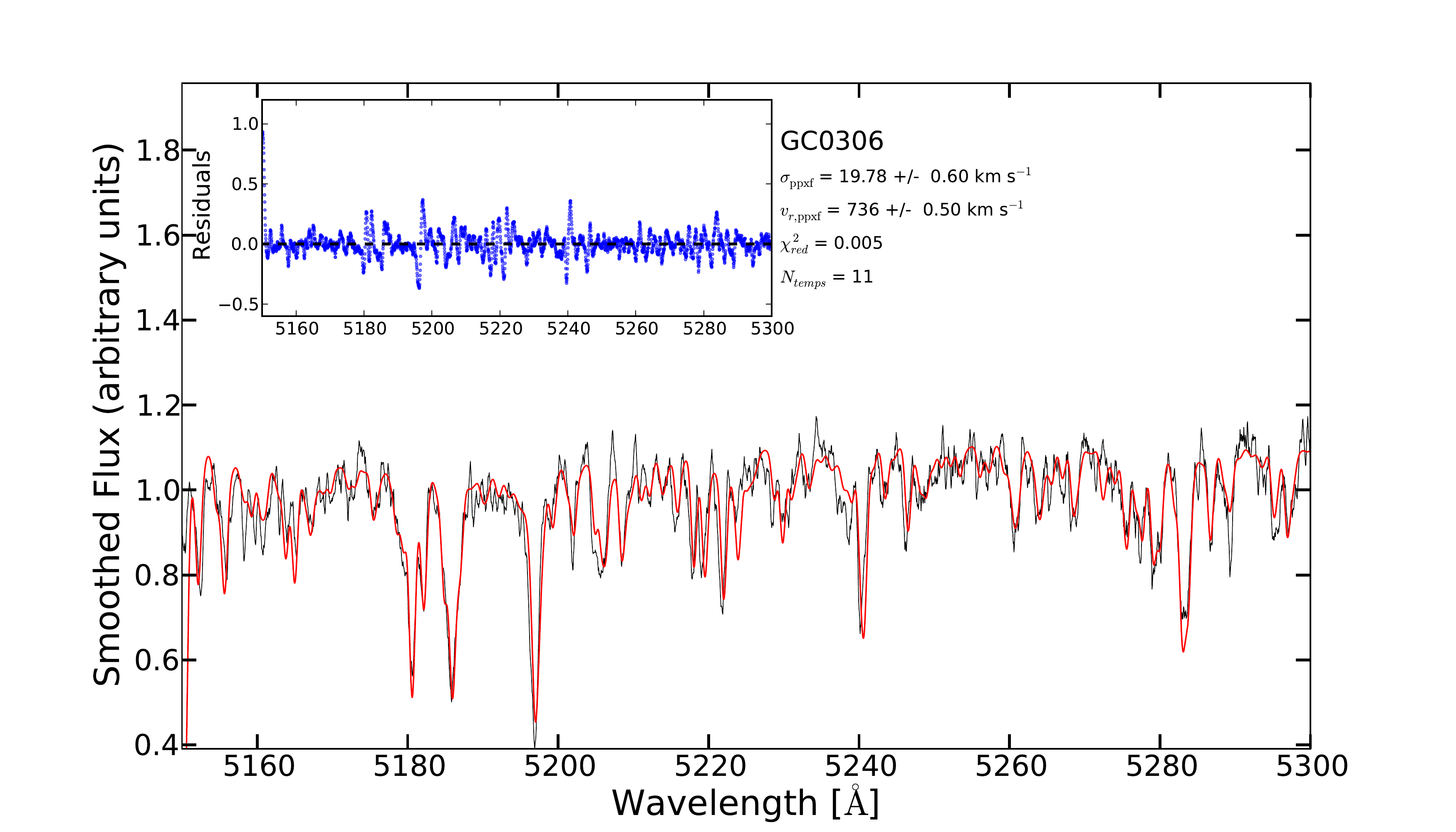}
\caption{Examples of the {\it ppxf} spectral fits (red lines) from a low-quality (top) to a high-quality spectrum (bottom). The reduced spectra (black lines) have been smoothed with a box size of 8 pixels, representative of the instrumental FWHM of GIRAFFE. The inset blue plots show the residuals, beside which we list some output information for each target. From top to bottom this information is the GC identification, our adopted $\sigma_{\rm ppxf}$, $v_{r,{\rm ppxf}}$, reduced $\chi^2$, and the number of template spectra used in their calculation.}
\label{fig:ppxf_fits}
\end{figure}

To account for any sensitivity that {\it ppxf} may have to the initial guess for $\sigma$ which may, for example, result in spurious solutions, we varied the input $\sigma$ between 5 and 100 km s$^{-1}$ in steps of 1 km s$^{-1}$ for each spectrum. For each fit we used the entire $5\,152\la\lambda/{\rm \AA}\la5\,365$ range, including the Mg$b$ and Fe features and all less prominent lines that are present.~All GIRAFFE spectra have a velocity scale of 2.87 km\,s$^{-1}$, and we found an additive 4$^{\rm th}$ degree polynomial adequate for the purpose of estimating the continua. After varying our initial $\sigma$ guesses, the adopted $v_{r,{\rm ppxf}}$ and $\sigma_{\rm ppxf}$ values correspond to the mean of the output sets of $v_{r,{\rm ppxf}}$ and $\sigma_{\rm ppxf}$ after being $\kappa\sigma$-clipped to remove outliers.~The {\it ppxf} errors were taken as the mean of the output {\it ppxf} errors added in quadrature to one standard deviation of the $\kappa\sigma$-clipped results.~While it is too cumbersome to present all of the {\it ppxf} fits in the present work, we direct the reader to the online-only appendix\footnote{\url{http://cdsarc.u-strasbg.fr/viz-bin/Cat/}} to access the full suite of spectral fits for all targets in this sample and present a representative sample in Figure~\ref{fig:ppxf_fits}.

\subsection{Template Library}
\label{sec:templates}
The {\it ppxf} code measurements rely on a library of template stellar spectra capable of accurately replicating an integrated light spectrum when used in combination and Doppler shifted to account for velocity gradients within a stellar population.~We used a library of high-resolution synthetic spectra from the PHOENIX\footnote{\url{http://phoenix.astro.physik.uni-goettingen.de/}} collaboration \citep{hus13}.~A synthetic spectral library was chosen over observed templates because current high resolution observed spectral libraries do not cover the wide range of stellar parameters that comprise complex stellar populations.~We therefore used a library of 1\,100 PHOENIX spectra covering the stellar parameter ranges: $0.5\!\leq\!\log g\!\leq\!4.0$, $3\,800\!\leq\!T_{\rm eff}\!\leq\!6\,000~{\rm K}$, $-4.0\!\leq\![{\rm Fe/H}]\!\leq\!+1.0$\,dex, and $-0.2\!\leq\![\alpha{\rm /Fe}]\!\leq+1.2$\,dex. For ${\rm [Fe/H]}\!=\!-4.0$ and ${\rm [Fe/H]}\!=\!+1.0$. The PHOENIX spectra are available only for $[\alpha/{\rm Fe}]\!=\!0.0$\,dex; however, we do not expect that this limitation affects our results significantly.

\subsection{Comparison with Previous Results}
\label{sec:tay10_comp}
\begin{figure}[t]
\centering
\includegraphics[width=8.9cm, bb=0 0 700 700]{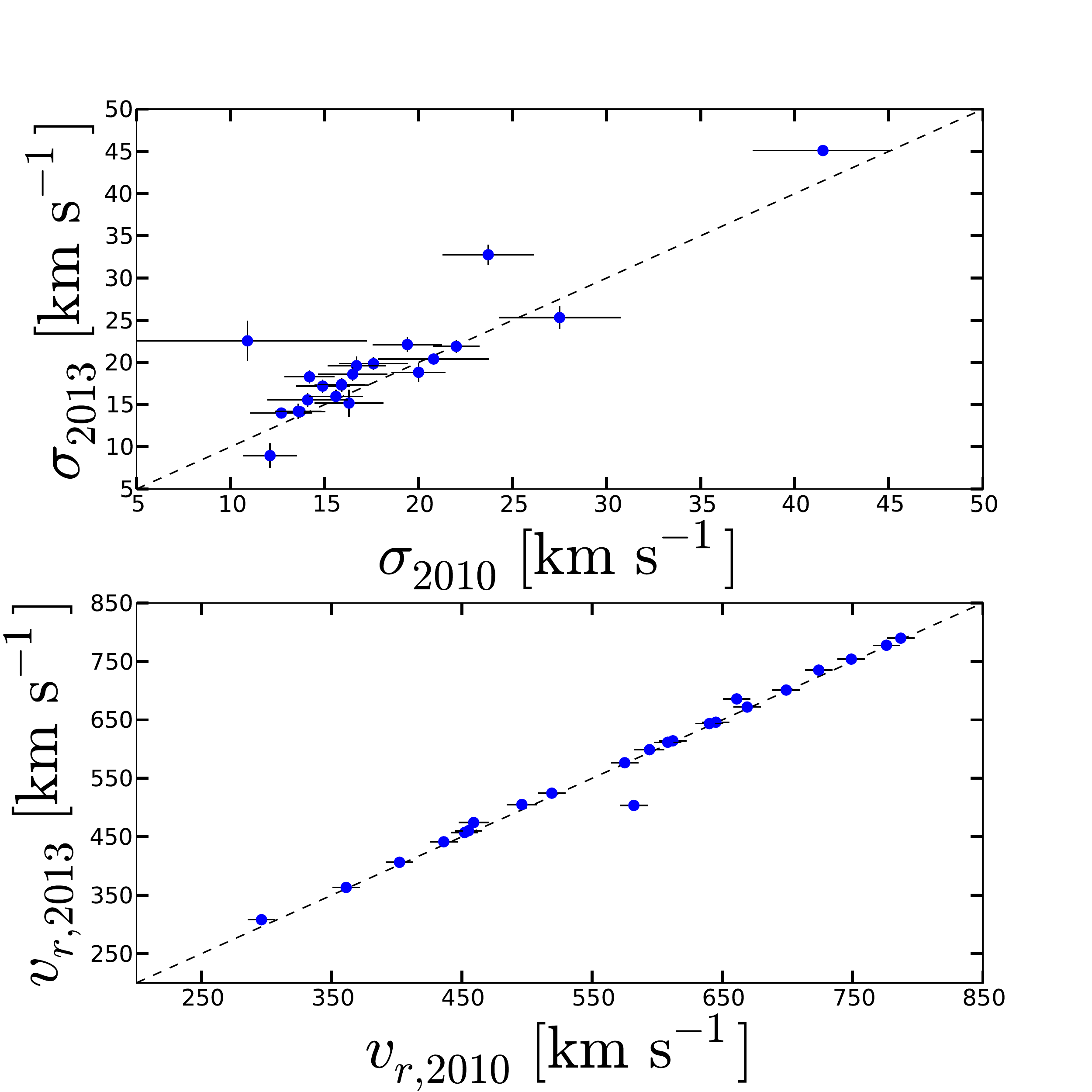}
\caption{Comparison of the {\it ppxf} accuracy using the GC sample of \citet{tay10} when the spectral range is restricted to that of the GIRAFFE spectra used in this work.~The upper panel shows the $\sigma$ comparison and the lower panel shows the same for $v_r$. In both panels, the \citet{tay10} results are shown along the x-axes and the new results are shown along the y-axes. Unity relations are indicated as dashed lines.}
\label{fig:tay10comp}
\end{figure}

The accuracy of {\it ppxf} when applied to the restricted $\sim\!200\,{\rm \AA}$ GIRAFFE wavelength coverage was tested by using the GC spectra of \cite{tay10} as a comparison sample since they also derived LOSVD estimates in the same manner, but with much wider spectral coverage.~To compare directly, the \citeauthor{tay10}~spectra were constrained to the GIRAFFE wavelength range, and $\sigma_{\rm ppxf}$ estimates were obtained as described in the following.~The input values for $v_r$ and $\sigma$ were fixed at those determined by \citeauthor{tay10}, and the input $\sigma$ were varied around the known values by $\pm10$\,km\,s$^{-1}$ in steps of 1 km\,s$^{-1}$.~The output values for $\sigma$ and $v_r$ are averaged and plotted as a function of input values in Figure~\ref{fig:tay10comp} for comparison.

The top panel of Figure~\ref{fig:tay10comp} shows that the agreement for $\sigma_{\rm ppxf}$ is generally good, with most of the GCs clustering around the unity relation.~The results of one GC (GC\,0382) are not shown, as our new LOSVD value of 178.77$\pm$7.58 km\,s$^{-1}$ is unlikely to be reliable considering that it implies a dynamical mass of $\sim\!10^9$\,$M_\odot$ within a half-light radius of $\sim\!2$\,pc.~Given that GC\,0382 has three independent $v_r$ measurements confirming it to be a member of \cena\ \citep[][and the present work]{woo10a}, and thus not a fore/background source, we adopt \citeauthor{tay10}'s $\sigma_{\rm ppxf}=14.3\pm3.2$ km s$^{-1}$ for the rest of the analysis.~The few other outliers in Figure~\ref{fig:tay10comp} correspond to the former study's most uncertain GCs, so we prefer our $\sigma_{\rm ppxf}$ estimates since our template library has a significantly wider range of stellar parameters and much higher S/N ratio over the wavelength range used to estimate $v_{r,{\rm ppxf}}$ and $\sigma_{\rm ppxf}$.~Meanwhile, the bottom panel shows that the agreement in $v_{r,{\rm ppxf}}$ is excellent, with the scatter around the unity relation being consistent with the measurement uncertainties.~The significant outlier corresponds to GC\,0382, which we consider to be unreliable and defer to any previously derived $v_r$ estimates in the literature.

Our new $v_{r,{\rm ppxf}}$ estimates are listed in column four of Table~\ref{tbl:kinematics}, compared to literature values listed in column three.~We note that GC\,0218, GC\,0219, GC\,0228 have $v_r$ measured for the first time ($528\pm1.90\,{\rm km\,s}^{-1}$, $661\pm2.70\,{\rm km\,s}^{-1}$ and $478\pm19.40\,{\rm km\,s}^{-1}$, respectively), all consistent with the 541\,km\,s$^{-1}$ systemic velocity of \cena\ \citep{woo10a}.~There were 15 GCs for which {\it ppxf} was unable to provide $v_r$ estimates, including GC\,0261 and GC\,0315, leaving them still as $v_r$ unconfirmed members of \cena.~Thus, Table~\ref{tbl:kinematics} lists new accurate $v_r$ estimates for 125 \cena\ GCs, including three first-time measurements. 

\begin{figure}[t]
\centering
\includegraphics[width=9cm, bb=10 0 834 768]{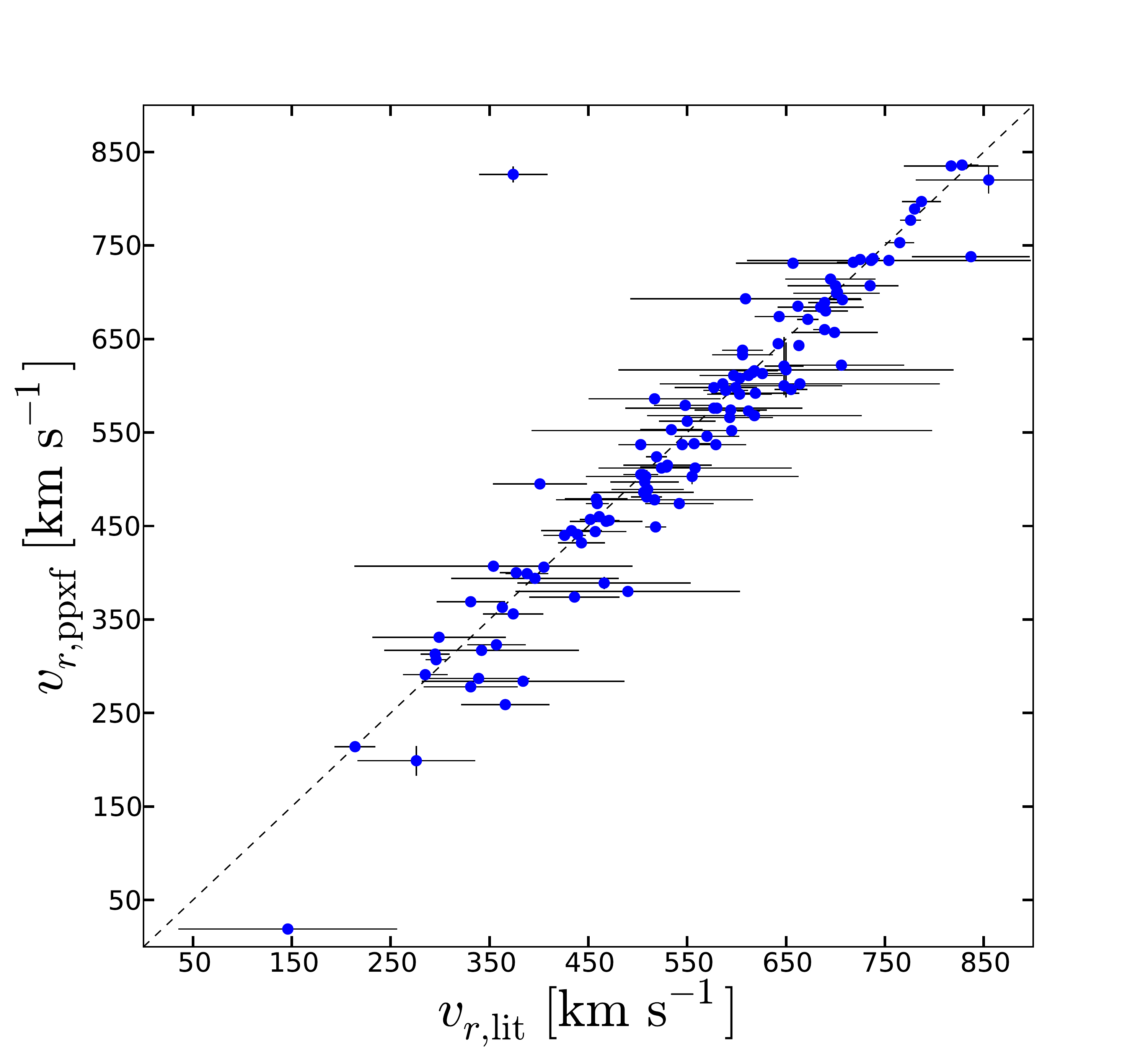}
\caption{Comparison between the radial velocities newly estimated by {\it ppxf}, $v_{r,{\rm ppxf}}$ to those previously measured from the literature, $v_{r,{\rm lit}}$.~The comparison is generally good, as indicated by the dashed line showing the unity relation.~However, our measurements are in most cases of significantly higher quality, as illustrated by the $v_{r,{\rm ppxf}}$ error bars, which for our dataset are often on the order of or smaller than the symbol size.}
\label{fig:vrcomp}
\end{figure}

We compare the new $v_{r,{\rm ppxf}}$ estimates to literature radial velocities ($v_{r,{\rm lit}}$) in Figure~\ref{fig:vrcomp} where $v_{r,{\rm lit}}$ and $v_{r,{\rm ppxf}}$ are shown along the $x$- and $y$-axes, respectively.~The \cite{woo07} and \cite{woo10a} catalogues provide the most comprehensive collections of \cena\ GC radial velocities to date.~For the comparison we adopt the weighted-average values listed in \cite{woo10a} where possible.~If there exists only a single \cite{woo10a} value measured from an individual spectrum, then we adopt the \cite{woo07} estimates, unless the former agrees significantly better with our new measurements.~Figure~\ref{fig:vrcomp} shows generally good agreement within the literature uncertainties, with a single notable exception being GC\,0095. For this GC, we prefer the literature measurement of $v_{r,{\rm lit}}=374\pm34\,{\rm km}\,{\rm s}^{-1}$ over our $\sigma_{\rm ppxf}=826\pm7.8\,{\rm km}\,{\rm s}^{-1}$ because visual inspection of the corresponding Doppler shifts shows better agreement with the laboratory wavelengths of the spectral absorption features when using the former value. Given this discrepancy, and the resulting uncertainty of the derived $\sigma_{\rm ppxf}$, we drop this object from the analysis. In any other case where $v_r$ are discrepant we prefer our $v_{r,{\rm ppxf}}$ due to smaller uncertainties.

Despite obtaining reliable $v_r$ estimates for almost all of our targets, there were 25 GCs for which we consider $\sigma_{\rm ppxf}$ to be unreliable either due to uncomfortably large error bars, or simply a failure to derive an estimate at all.~We therefore drop these targets from the subsequent analysis and carry on with the remaining 115 new $\sigma_{\rm ppxf}$ estimates, including the re-analyzed \citeauthor{tay10} GCs.

\subsection{Structural Parameters}
\label{sec:struc_par}
We took 2D projected half-light radii, $r_h$, and concentration parameters, $c$, from the three sources as listed in Table~\ref{tbl:structural}.~The majority of the values listed for $r_h$ and $c$ are, where available, from Jord\'an et al.~(2015, {\it in prep.}), based on {\it HST} data, and are otherwise taken from Gomez et al.~(2015, {\it in prep.}) based on {\it IMACS} data taken under exceptional seeing ($\lesssim\!0.5\arcsec$).~Despite the sub-arcsecond seeing conditions, the marginally resolved nature of most GCs did not allow for accurate $c$ estimates by Gomez et al., so for many, a typical value of 1.48 was assigned. Moreover, three extremely small {\it IMACS}-based $r_h$ estimates exist where there is no {\it HST} imaging available; however, these GCs (GC\,0085, 0333, 0429) are not resolved in the images and thus we drop them from the analysis and continue with the remaining 112 CSSs.~The re-analyzed clusters of \cite{tay10} use the same parameters as in that paper, namely those derived in \cite{har02a}.~For the latter GCs, the $r_h$ errors listed are adopted from the same paper. At the time of writing, no errors were available for the Gomez et al. and Jord\'an et al. sizes, so for these clusters we adopt values of 0.43 pc, which is the average size error of the \cite{tay10} sample, and while representative for the new {\it HST} measurements, may underestimate the {\it IMACS} errors.~Following \cite{har02a}, we assume 0.15 for the errors on all $c$, noting that these probably underestimate the true {\it IMACS}-based measurement errors as well.~In any case, although assigning error estimates in this manner is not optimal, we do not expect it to affect our main results significantly given the dominance of $\sigma$ on the uncertainties of our dynamical mass estimates (see \S~\ref{sec:masses}).

\begin{deluxetable*}{lccccccccc}
\tabletypesize{\scriptsize}
\tablecaption{Star Cluster Kinematics\label{tbl:kinematics}}
\tablewidth{0pt}
\tablehead{
\colhead{ID} & \colhead{$R_{\rm gc}$} & \colhead{$v_{r,{\rm lit}}$} & \colhead{$v_{r,{\rm ppxf}}$} & \colhead{$\sigma_{\rm ppxf}$} & \colhead{$\sigma_{m,ap}$} & \colhead{$\sigma_{m,gl}$} & \colhead{$\sigma_0$} & \colhead{$\sigma_{1/2}$} & \colhead{Ref.} \\
\colhead{} & \colhead{[arcmin]} & \colhead{[km s$^{-1}$]} & \colhead{[km s$^{-1}$]} & \colhead{[km s$^{-1}$]} & \colhead{[km s$^{-1}$]} & \colhead{[km s$^{-1}$]} & \colhead{[km s$^{-1}$]} & \colhead{[km s$^{-1}$]} & \colhead{}\\
\colhead{(1)} & \colhead{(2)} & \colhead{(3)} & \colhead{(4)} & \colhead{(5)} & \colhead{(6)} & \colhead{(7)} & \colhead{(8)} & \colhead{(9)} & \colhead{(10)}
}
\startdata
GC\,0028 & 11.30 	& 558$\pm$97 				& 512$\pm$1.80 	& 8.95$\pm$2.10 	& 8.97$^{+2.06}_{-2.11}$		&	8.60$^{+1.80}_{-1.92}$	&	 9.87$^{+2.26}_{-2.25}$	&	9.52$^{+2.27}_{-2.25}$ 	&	1 \\
GC\,0031 & 10.63 	& 595$\pm$202 			& 552$\pm$2.80 	& 3.69$\pm$4.35 	& 3.67$^{+4.34}_{-3.65}$		&	3.54$^{+4.05}_{-3.53}$	&	4.12$^{+4.77}_{-4.11}$	&	3.94$^{+4.68}_{-3.93}$ 	&	1 \\
GC\,0048 & 11.36 	& 509$\pm$15 				& 481$\pm$1.10 	& 7.61$\pm$1.40 	& 7.62$^{+1.38}_{-1.38}$		&	7.08$^{+1.13}_{-1.18}$	&	 8.25$^{+1.54}_{-1.52}$	&	7.93$^{+1.48}_{-1.45}$ 	&	1 \\
GC\,0050 & 8.03 	& 718$\pm$16 				& 732$\pm$0.70 	& 10.09$\pm$0.90 	& 10.01$^{+0.98}_{-0.82}$	&	9.01$^{+0.71}_{-0.61}$	&	10.65$^{+0.82}_{-0.96}$	&	10.08$^{+0.92}_{-0.84}$ 	&	1 \\
GC\,0052 & 7.89 	& 276$\pm$59 				& 199$\pm$15.25 	& 1.09$\pm$51.85 	& \nodata					& \nodata 						& \nodata				&	\nodata				&	2 \\
GC\,0053 & 7.75 	& 503$\pm$17 				& 505$\pm$1.00 	& 10.50$\pm$1.20 	& 10.55$^{+1.10}_{-1.28}$	&	9.95$^{+0.83}_{-1.03}$	&	11.91$^{+1.25}_{-1.42}$	&	11.31$^{+1.29}_{-1.41}$ 	&	1 \\
GC\,0054 & 8.12 	& 736$\pm$58			 	& 734$\pm$1.50 	& 9.31$\pm$1.70 	& 9.30$^{+1.72}_{-1.68}$		&	8.16$^{+1.36}_{-1.38}$	&	 9.60$^{+1.84}_{-1.83}$	&	9.23$^{+1.63}_{-1.75}$	&	3 \\
GC\,0058 & 8.06 	& 685$\pm$43				& 684$\pm$0.80 	& 7.60$\pm$1.10 	& 7.61$^{+1.06}_{-1.11}$		&	7.47$^{+0.92}_{-1.03}$	&	 8.50$^{+1.32}_{-1.22}$	&	8.24$^{+1.23}_{-1.31}$ 	&	2 \\
GC\,0064 & 9.42 	& 594$\pm$36 				& 574$\pm$2.90 	& 16.29$\pm$2.90 	& 16.27$^{+2.89}_{-2.89}$	&	15.29$^{+2.40}_{-2.50}$	&	17.79$^{+3.35}_{-2.95}$	&	17.08$^{+3.14}_{-3.06}$ 	&	1 \\
GC\,0065 & 6.92 	& 331$\pm$47			 	& 278$\pm$1.30 	& 9.78$\pm$1.62 	& 9.78$^{+1.58}_{-1.64}$		&	9.13$^{+1.26}_{-1.37}$	&	11.25$^{+1.76}_{-1.96}$	&	10.55$^{+1.71}_{-1.78}$	&	2
\enddata
\tablecomments{Kinematical data for the NGC 5128 star clusters. Cols.~1 and 2 list the cluster IDs and projected galacto-centric radii respectively, cols.~3 and 4 list radial velocities, and cols.~5-9 list $\sigma$ measured with {\it ppxf} and values which have been aperture-corrected to various cluster radii (see \S~\ref{sec:apcorr} for details).~Where available, all $v_{r,{\rm lit}}$ values are taken from \cite{woo10a} corresponding to their ``mean'' values otherwise we adopt the best matches from either \cite{woo07} or the estimates from \cite{woo10a} that are based on individual spectra. Table~\ref{tbl:kinematics} is published in its entirety in the electronic edition of the {\it Astrophysical Journal}. A portion is shown here for guidance regarding its form and content.\\
References for $v_{r,{\rm lit}}$. (1) \cite{woo10a}; (2) \cite{woo07}, their mean; (3) \cite{woo07}, LDSS2; (4) \cite{woo07}, VIMOS; (5) \cite{woo07}, Hydra.}
\end{deluxetable*}

\subsection{Aperture Corrections}
\label{sec:apcorr}
The $\sigma_{\rm ppxf}$ values listed in Table~\ref{tbl:kinematics}, while generally accurate, are not appropriate to use when estimating dynamical masses.~Several effects, both observational (seeing, target distance, etc.) and instrumental (spectral/spatial resolution, sampling, etc.), may conspire to affect how representative the light entering a given fibre aperture may be of objects similar to massive GCs and UCDs \citep{mie08a}.~In our case the 1.2\arcsec\ diameters of the FLAMES fibres correspond to $\sim\!22$\,pc at the distance of \cena, so contributions from stars outside of the core region may skew the $\sigma_{\rm ppxf}$ measurements to lower values compared to $\sigma$ estimates corresponding to smaller radii.~Here we describe our approach to correct our measured $\sigma_{\rm ppxf}$ estimates to values representing both the GC core regions ($\sigma_0$) and $\sigma$ values within the GC half-light radius ($\sigma_{1/2}$).

We used the cluster modeling code of \cite{hil07}, described in detail in \cite{mie08a}, to correct for any aperture effects and determine estimates of $\sigma_0$ and $\sigma_{1/2}$ based on our measured $\sigma_{\rm ppxf}$.~This code uses the basic structural data (in our case $r_h$ and $c$) that defines a cluster's light-profile to generate a 3D \cite{kin66} stellar density profile from which an N-body representation of the cluster is created in 6D (position, velocity) space.~Each simulated particle is convolved with a Gaussian corresponding to the true seeing FWHM (see \S~\ref{sec:obs}) and a light profile is generated from which the velocity dispersion profile can be obtained.

Using this code, we modeled each of our clusters with $10^5$ particles and binned them radially in groups of 10$^3$.~The 3D velocity information of each subgroup was used to derive $\sigma$ profiles according to $\langle v\rangle^2_{\rm 3D}=3\sigma^2$ where $\langle v\rangle^2_{\rm 3D}$ is the square of the mean of the 3D velocities.~To account for the inherent stochasticity of the modeling, the median of the inner-most five subgroups, or 5\% of the modeled stellar population, was adopted as $\sigma_0$, while all particles inwards of $r_h$ were used to calculate $\sigma_{1/2}$ for each GC.~This process was repeated three times per GC.~The first set of models used the measured $\sigma_{\rm ppxf}$, $r_h$, and $c$ as inputs to provide our adopted $\sigma_0$ and $\sigma_{1/2}$, while for the other two iterations we added or subtracted the errors for the three quantities in order to maximize or minimize the modeled $\sigma$ estimates, respectively.~We then adopted the differences between the upper/lower bounds and the output $\sigma$ values as the corresponding errors.~Table~\ref{tbl:kinematics} lists the resulting $\sigma_{0}$ and $\sigma_{1/2}$ estimates including our uncertainties (columns 8 and 9), alongside the model velocity dispersions corresponding to the FLAMES apertures ($\sigma_{m,ap}$) and the global values ($\sigma_{m,gl}$).~The accuracy of the code is verified by the very good agreement between the predicted $\sigma_{\rm ppxf}$ and measured $\sigma_{m,ap}$ at the fibre aperture size.

\subsection{Star Cluster Masses and Mass-to-Light Ratios}
\label{sec:masses}
One of the most direct methods to estimate the dynamical mass (${\cal{M}}_{\rm dyn}$) of a single-component compact stellar system is by the use of the scalar virial theorem \citep[e.g.][]{bin08} of the form originally derived by \cite{spi69},
\begin{equation}
\label{eq:mdyn}
{\cal{M}}_{\rm dyn}\simeq2.5\frac{3\sigma_0^2r_h}{G}\simeq1743\left(\frac{\sigma_0^2}{{\rm km}^2~{\rm s}^{-2}}\right)\left(\frac{r_h}{{\rm pc}}\right){\rm M}_\odot
\end{equation}
if one assumes a dynamically relaxed cluster, sphericity, and isotropic stellar orbits.~While this is among the most commonly used dynamical mass estimators, it has been shown that the ``half-mass'' (${\cal{M}}_{1/2}$) or in other words the dynamical mass corresponding to that contained within the 2D projected half-light radius is more robust against stellar velocity dispersion anisotropy.~This feature makes ${\cal{M}}_{1/2}$ an overall more robust mass estimator for dispersion supported systems.~We estimate ${\cal{M}}_{1/2}$ via the form derived by \cite{wol10},
\begin{equation}
\label{eq:halfmass}
{\cal{M}}_{1/2}=4\frac{\langle\sigma_{\rm los}^2\rangle r_h}{G}\simeq930\left(\frac{\langle\sigma_{\rm los}^2\rangle}{{\rm km}^2~{\rm s}^{-2}}\right)\left(\frac{r_h}{{\rm pc}}\right){\rm M}_\odot
\end{equation}
where $\sigma_{\rm los}$ is the luminosity-weighted LOSVD, in our case aperture corrected to $\sigma_{1/2}$.

Applying Equation~\ref{eq:halfmass} to all the GCs with available $\sigma_{1/2}$, $r_h$, and $c$ provides ${\cal{M}}_{1/2}$ estimates for a total of 112 of \cena\ star clusters, 89 of which are first-time measurements, in particular at faint absolute luminosities (see Section~\ref{sec:md_v_ml} and Figure~\ref{fig:mag_mld_size}).~We find in our star cluster sample ${\cal{M}}_{1/2}$ estimates ranging from the low-mass end, ${\cal{M}}_{1/2,{\rm min}}\!=\!3.7^{+8.9}_{-3.7}\cdot10^4\,M_\odot$ (GC\,0031), to the highest-mass object, GC\,0365, with ${\cal{M}}_{1/2,{\rm max}}\!=\!7.41^{+0.51}_{-0.63}\cdot10^6\,M_\odot$, with a sample median ${\cal{M}}_{1/2}$ of 3.47$\cdot10^5\,M_\odot$.~By assuming that mass follows light, these masses translate into total mass estimates of ${\cal{M}}_{{\rm tot},{\rm min}}\!=\!7.4^{+12.6}_{-7.4}\cdot10^4\,M_\odot$ representative of the lower range of GC masses, ${\cal{M}}_{{\rm tot},{\rm max}}\!=\!1.48\!^{+0.72}_{-0.89}\cdot10^7\,M_\odot$ consistent with UCD masses, and a median ${\cal{M}}_{\rm tot}\!=6.94\!\cdot10^5\,M_\odot$.

\begin{deluxetable*}{lccccccc}
\tabletypesize{\scriptsize}
\tablecaption{Star Cluster Structural Parameters\label{tbl:structural}}
\tablewidth{0pt}
\tablehead{
\colhead{ID} & \colhead{${\cal{M}}_{1/2}$} & \colhead{${\cal{M}}_{\rm dyn}$} & \colhead{$\Upsilon_{1/2}$} & \colhead{$\Upsilon_{\rm dyn}$} & \colhead{$r_{h}$} & \colhead{$c$} & \colhead{Ref.} \\
\colhead{} & \colhead{[$M_{\odot}$]} & \colhead{[$M_{\odot}$]} & \colhead{[$M_{\odot}/L_{\odot}$]} & \colhead{[$M_{\odot}/L_{\odot}$]} & \colhead{[pc]}  & \colhead{} \\
\colhead{(1)} & \colhead{(2)} & \colhead{(3)} & \colhead{(4)} & \colhead{(5)} & \colhead{(6)} & \colhead{(7)} & \colhead{(8)}
}
\startdata
GC0028	   & 0.25$^{+0.12}_{-0.12}$	&        0.50$^{+0.24}_{-0.24}$	&	 3.42$^{+1.71}_{-1.70}$	&	 3.45$^{+1.66}_{-1.66}$	& 2.96$\pm$0.43 & 1.48 	&	1 \\
GC0031	   & 0.04$^{+ 0.09}_{- 0.04}$	&	 0.08$^{+0.18}_{-0.08}$	&	 0.49$^{+1.16}_{-0.48}$	&	 0.50$^{+1.16}_{-0.49}$ 	& 2.58$\pm$0.43 & 1.48 	&	1 \\
GC0048	   & 0.24$^{+0.09}_{- 0.09}$	&	 0.48$^{+0.19}_{-0.19}$	&	 2.50$^{+0.98}_{-0.96}$	&	 2.54$^{+0.99}_{-0.98}$ 	& 4.08$\pm$0.43 & 1.65 	&	2 \\
GC0050	   & 0.66$^{+0.13}_{- 0.12}$	&	 1.39$^{+0.23}_{-0.26}$	&	 3.43$^{+0.68}_{-0.63}$	&	 3.59$^{+0.62}_{-0.70}$ 	& 7.03$\pm$0.43 & 1.56 	&	2 \\
GC0052	   & \nodata				& 	\nodata				& 	\nodata				 & 	\nodata			   	& 1.02$\pm$0.43 & 1.48 	&	1 \\
GC0053	   & 0.33$^{+0.09}_{- 0.10}$	&	 0.68$^{+0.18}_{-0.19}$	&	 3.61$^{+1.02}_{-1.08}$	&	 3.76$^{+1.00}_{-1.09}$ 	& 2.74$\pm$0.43 & 1.83 	&	2 \\
GC0054	   & 0.71$^{+0.25}_{- 0.27}$	&	 1.45$^{+0.56}_{-0.56}$	&	 4.04$^{+1.45}_{-1.56}$	&	 4.10$^{+1.59}_{-1.59}$ 	& 9.01$\pm$0.43 & 1.65 	&	2 \\
GC0058	   & 0.12$^{+0.05}_{- 0.05}$	&	 0.24$^{+0.09}_{-0.09}$	&	 1.14$^{+0.43}_{-0.45}$	&	 1.14$^{+0.44}_{-0.42}$ 	& 1.89$\pm$0.43 & 1.43 	&	1 \\
GC0064	   & 1.02$^{+0.39}_{- 0.39}$	&	 2.08$^{+0.82}_{-0.73}$	&	18.65$^{+7.24}_{-7.07}$	&	18.97$^{+7.52}_{-6.72}$ 	& 3.77$\pm$0.43 & 1.57 	&	1 \\
GC0065	   & 0.28$^{+0.10}_{- 0.11}$	&	 0.60$^{+0.21}_{-0.23}$	&	 2.23	$^{+0.81}_{-0.84}$	&	 2.38$^{+0.84}_{-0.92}$ 	& 2.71$\pm$0.43 & 1.99	&	1
\enddata
\tablecomments{Structural parameters and dynamical masses of our GC sample. GC\,0082, GC\,0107, GC\,0214, GC\,0219, GC\,0236, GC\,0262, GC\,0274, GC\,0417, GC\,0418, GC\,0420, GC\,0439, GC\,0435, and GC\,0437 are based on de-reddened $r'$ band data. Table~\ref{tbl:structural} is published in its entirety in the electronic edition of the {\it Astrophysical Journal}. A portion is shown here for guidance regarding its form and content.\\
References for $r_h$ and $c$: (1) Gomez et al. (2015, {\it in prep.)}; (2) Jord\'an et al. (2015, {\it in prep.}); (3) \cite{har02a}.}
\end{deluxetable*}

Alternatively, we use $\sigma_0$ to estimate ${\cal{M}}_{\rm dyn}$ via Equation~\ref{eq:mdyn}. By doing this, we lose the benefit of isotropy independence, but gain freedom from the underlying assumption that mass follows light.~Encouragingly, we find very similar masses using either estimator at the low-mass tail of the GC mass distribution, with ${\cal{M}}_{{\rm dyn},{\rm min}}\!=\!7.6^{+17.7}_{-7.6}\cdot10^4\,M_\odot$, and at the highest mass with ${\cal{M}}_{{\rm dyn},{\rm max}}\!=\!1.53^{+0.12}_{-0.15}\cdot10^7\,M_\odot$ and a median ${\cal{M}}_{\rm dyn}$ of $7.04\cdot10^5\,M_\odot$.~Interestingly, for total masses higher than $\sim\!1.0\cdot10^6\,M_\odot$ we find that the relative difference between the estimators becomes more significant for certain clusters.

\begin{figure}[b]
\centering
 \includegraphics[width=8.9cm, bb=0 0 970 925]{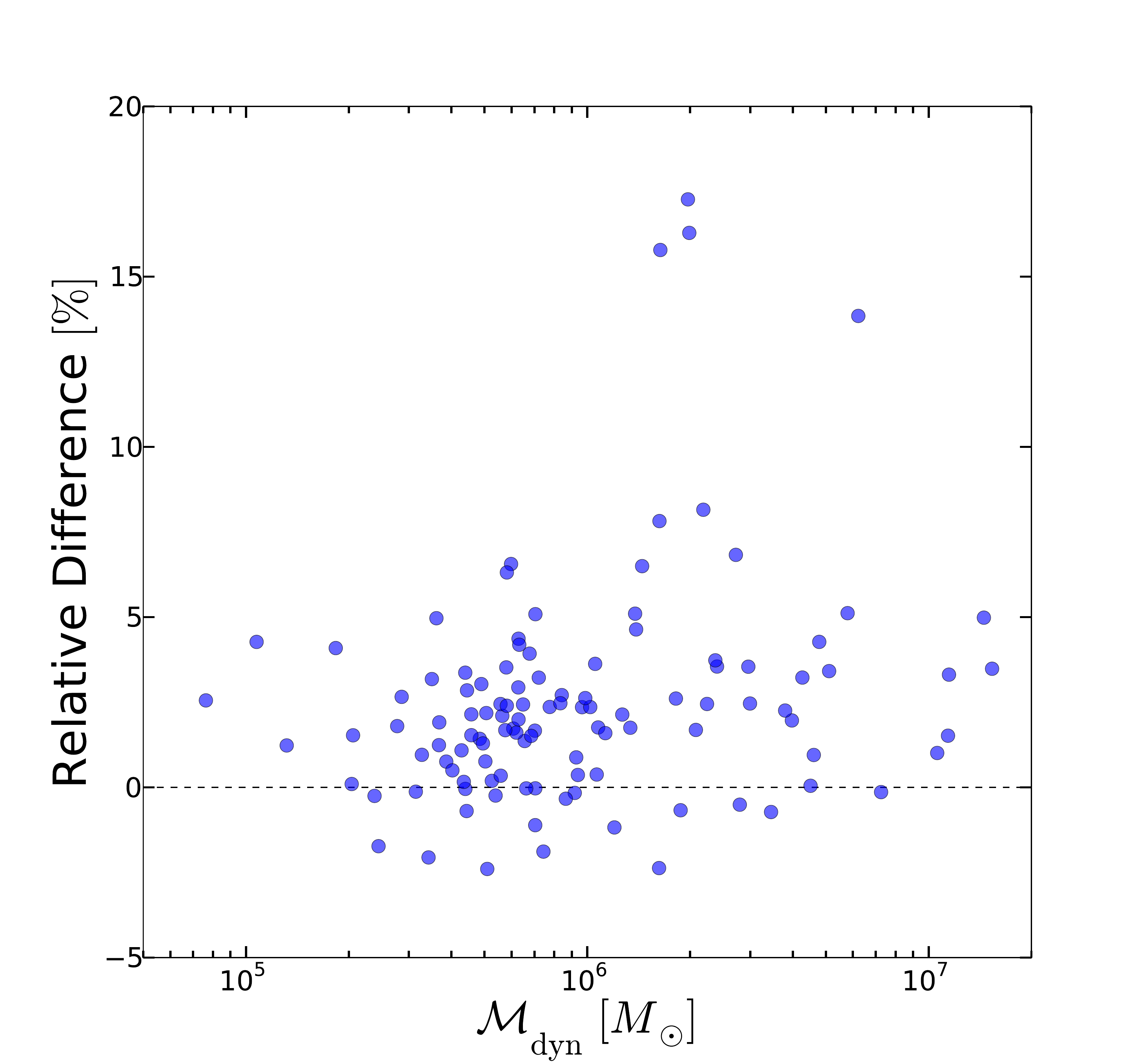}
 \caption{Relative difference between the traditionally used dynamical mass ${\cal{M}}_{\rm dyn}$ and twice the half-mass, i.e.~${\cal{M}}_{\rm dyn}\!-\!2\!\cdot\!{\cal{M}}_{1/2}$, as a function of dynamical mass, ${\cal{M}}_{\rm dyn}$.~If mass follows light, then the total GC mass should be approximated by twice ${\cal{M}}_{1/2}$, hence, the difference being close to zero (dashed line).~Here, ${\cal{M}}_{\rm dyn}$ predicts somewhat higher masses than $2\!\cdot\!{\cal{M}}_{1/2}$ for GCs with ${\cal{M}}_{\rm dyn}\!\ga\!10^6\,M_\odot$, indicating that for these clusters, mass may not strictly follow light and either dark gravitating mass components or non-equilibrium configurations may become necessary to explain the discrepancy in dynamical mass estimates of these clusters.}
 \label{fig:md_v_mh}
\end{figure}

Figure~\ref{fig:md_v_mh} shows the relative difference between the total dynamical masses derived using $2\cdot {\cal{M}}_{1/2}$ and ${\cal{M}}_{\rm dyn}$ as a function of ${\cal{M}}_{\rm dyn}$ for each GC in our sample.~${\cal{M}}_{\rm dyn}$ essentially predicts similar (within $\sim1-5\%$) masses for ``normal'' GCs, ($\la\!10^6\,M_\odot$; typical of Local Group GCs).~Above this threshold the discrepancy between the two mass estimates becomes more pronounced for certain clusters, reaching up to $\sim18\%$ higher ${\cal{M}}_{\rm dyn}$.~Altogether, this comparison suggests that for typical GC masses, i.e.~in the range $\sim\!10^5\!-\!10^6\,M_\odot$, where cluster masses are completely dominated by baryonic material, ${\cal{M}}_{1/2}$ is the more robust measure of the total GC mass. Above this mass range, and outside of $r_h$ any kinematical tracers arising from, e.g.~non-equilibrium configurations or/and dark gravitating mass components, may introduce biases in the mass estimates.

We calculate the dynamical mass-to-light ratios evaluated within the half-light radius ($\Upsilon^{\rm 1/2}_V$) by dividing $2\!\cdot\!{\cal{M}}_{1/2}$ by the total, de-reddened $V$-band luminosity, calculated as,
\begin{equation}
L_V=10^{-0.4(V_0-(m-M)_0-M_{V,\odot})}
\end{equation}
where $M_{V,\sun}\!=\!4.83$ mag.~We also calculate dynamical mass-to-light ratios based on $\sigma_0$ ($\Upsilon_V^{\rm dyn}$) by dividing ${\cal{M}}_{\rm dyn}$ by $L_V$.~Most of our sample have apparent $V$-band magnitudes provided in the \cite{woo07} catalogue (see also references therein), for which we list the de-reddened values in Table~\ref{tbl:observations}.~We account for foreground reddening on an individual basis, with no attempt to correct for extinction internal to \cena, by using the {\it Galactic Extinction and Reddening Calculator}\footnote{\url{http://ned.ipac.caltech.edu/forms/calculator.html}} with the galactic reddening maps of \cite{sch11}.~Where $V$-band magnitudes are not available, we list de-reddened $r'$-band magnitudes from \cite{sin10}.~If no photometry in the previously mentioned filters is available, we base our $\Upsilon_V$ estimates on the $R$-band magnitudes from the acquisition images, assuming in all cases a conservative photometric error of 0.1 mag.

We find a large spread in the corresponding $\Upsilon^{1/2}_V$ ranging between 0.49\,$M_\odot L^{-1}_{V,\odot}$ for GC\,0031 up to 64.47\,$M_\odot L^{-1}_{V,\odot}$ for GC\,0225, with a sample median $\Upsilon^{1/2}_V$ of 3.33\,$M_\odot L^{-1}_{V,\odot}$.~Similarly, we find $0.50\leq\Upsilon_V^{\rm dyn}\leq66.61\,M_\odot L^{-1}_{V,\odot}$ with a slightly higher median value of 3.44\,$M_\odot L^{-1}_{V,\odot}$.~While these results are smaller by 0.47 and 0.36 $M_\odot L^{-1}_{V,\odot}$ than the median of 3.8\,$M_\odot L^{-1}_{V,\odot}$ found by \cite{tay10} for \cena\ GCs, they are notably higher than the median $\Upsilon_{V,{\rm dyn}}$ of $2.2\pm0.3$ for Milky Way (MW) GCs \citep{mcl00,mcl08} and $\Upsilon_{V,{\rm dyn}}\!=1.37\pm0.28$ for M31 GCs \citep[][]{str11}.~This result should perhaps not be too surprising as our sample is biased toward GCs at the bright end of the GC luminosity function (GCLF).~Thus, we are most likely not including many GCs with typical Local Group GC masses so that our sample is biased to GCs above the $\sim2\cdot10^6\,M_\odot$ threshold where $\Upsilon^{\rm dyn}_V$ begins to rise dramatically \citep[e.g.][]{has05, kis06, mie06, mie08a, tay10}.~Conversely, our sample includes not only those of \citeauthor{tay10}, but many fainter GCs, thus reaching well below the aforementioned threshold and biasing our medians toward slightly lower values compared to previous studies.

\begin{figure*}[t]
\centering
\includegraphics[width=0.97\linewidth, bb=0 0 800 710]{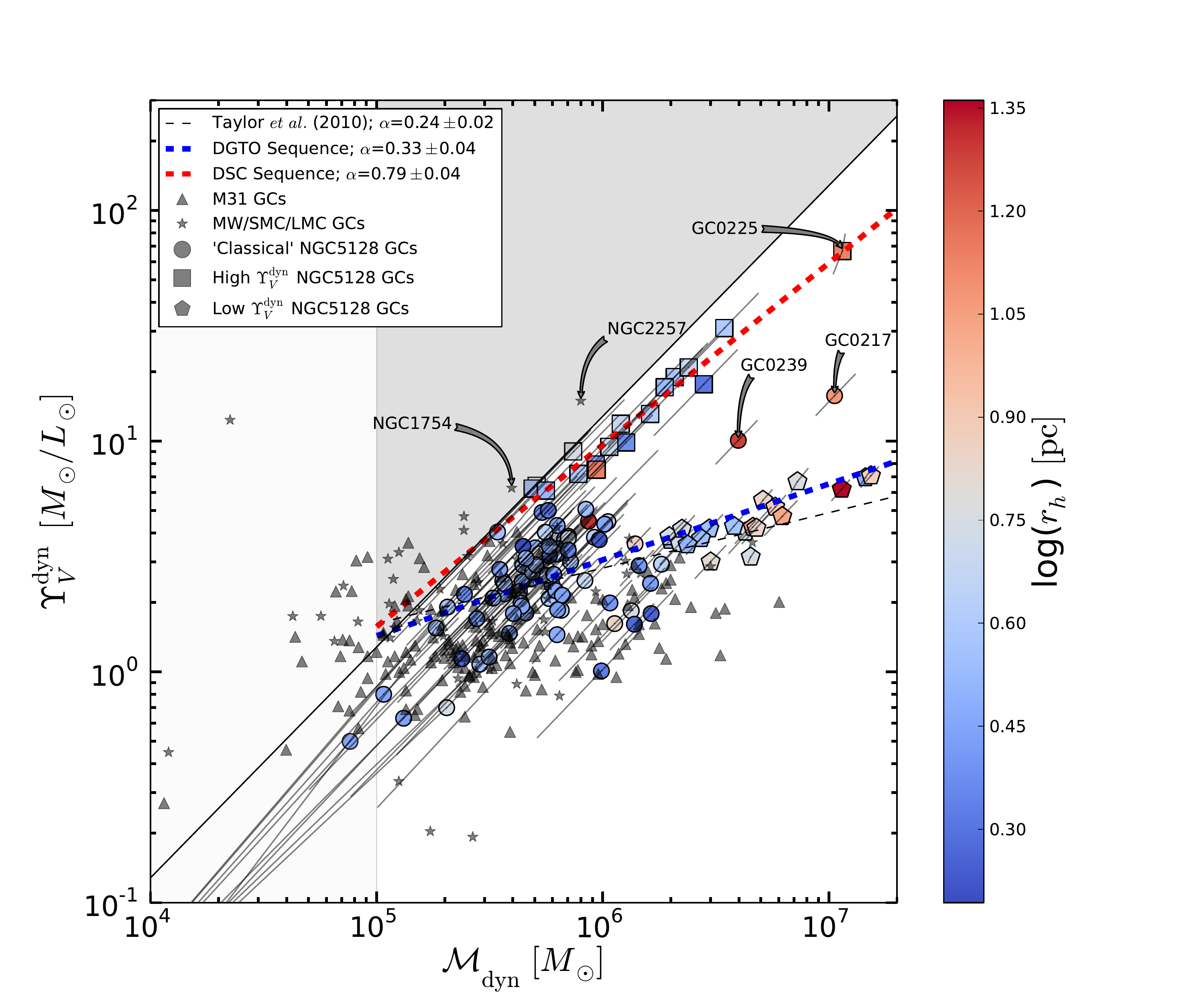}
\caption{$\Upsilon_V^{\rm dyn}$ as a function of ${\cal{M}}_{\rm dyn}$, with the color shading indicating the half-light radius of the clusters.~Three groups of GCs corresponding to ``classical'' GCs, and to the ``low-'' and  ``\hdyn '' sequence GCs are indicated by circular, pentagonal, and square symbols, respectively (see legend).~The empirical power-law relation of \citet{tay10} is shown by the thin dashed black line, while steeper power-law fits based on the two $\Upsilon_V^{\rm dyn}$ branches are shown by the colored thick dashed lines.~The steepest relation (red) is based only on the square points which have been isolated based on their faint absolute magnitudes ($M_V\!\ga\!-8.5$\,mag) and high mass-to-light ratios ($\Upsilon_V^{\rm dyn}\!\geq\!6.0\,M_\odot L_{V,\odot}^{-1}$). Meanwhile, the blue dashed line shows the results of fitting only the pentagonal points with $\Upsilon_V^{\rm dyn}\!\leq\!10.0\,M_\odot L_{V,\odot}^{-1}$ and ${\cal{M}}_{\rm dyn}\geq2\cdot10^6\,M_\odot$.~The solid black line and dark gray shading illustrate the observational limit to our data, while the light gray shading indicates an approximate lower mass limit for selected GCs in our spectroscopic sample.~Several interesting objects that are discussed in the text are labeled with their catalog numbers.~We plot corresponding measurements for Local Group GCs as grey stars \citep[MW/SMC/LMC; taken from][]{mcl05} and grey triangles \citep[M31; taken from][]{str11}.}
\label{fig:mass_mld_size}
\end{figure*}

\section{Discussion}
\label{sec:disc}
\begin{figure*}
\centering
 \includegraphics[width=0.97\linewidth, bb=0 0 800 710]{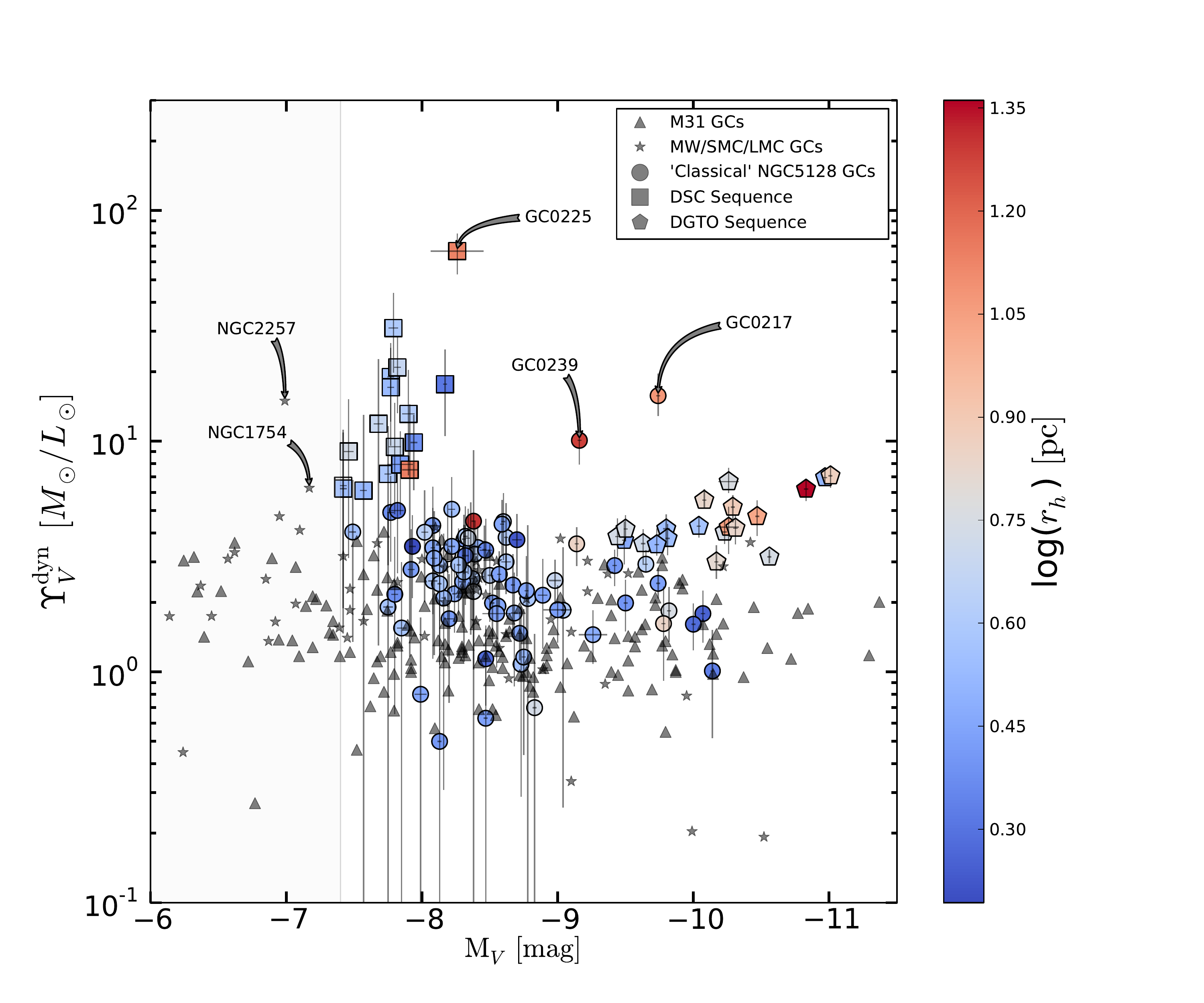}
 \caption{$\Upsilon_V^{\rm dyn}$ as a function of $M_V$, with cluster half-light radius indicated by the color bar.~Symbols are the same as in Fig.~\ref{fig:mass_mld_size}, with the same objects indicated. The vertical light-grey line and the shaded zone ($M_V\!\geq\!-7.4$ mag) indicate the photometric limit of our spectroscopic sample selection.}
 \label{fig:mag_mld_size}
\end{figure*}

Correlations between $\Upsilon_V^{\rm dyn}$, ${\cal M}_{\rm dyn}$, and absolute magnitude ($M_V$) can provide important information on the dynamic configuration and baryonic makeup of star clusters.~To investigate these relations, $\Upsilon_V^{\rm dyn}$ is shown as a function of ${\cal{M}}_{\rm dyn}$ and $M_V$ for each of our sample GCs in Figures~\ref{fig:mass_mld_size} and \ref{fig:mag_mld_size} respectively, with the color shading parametrizing GC $r_h$.

Given that the following discussion hinges strongly on the features seen in Figures \ref{fig:mass_mld_size} and \ref{fig:mag_mld_size} having astrophysical explanations, we first considered several systematic effects and performed corresponding tests to check whether these could bias our measurements and artificially generate the observed results.~The description of these tests, including detailed checks for data analysis biases, target confusion, correlations with galactocentric radius ($R_{\rm gc}$), and/or insufficient background light subtraction is provided in the Appendix.~In summary, none of the tested effects are likely to explain the observed features in Figure~\ref{fig:mass_mld_size} and \ref{fig:mag_mld_size}, and thus we consider astrophysical explanations in what follows.

\subsection{${\cal M}_{\rm dyn}$ vs. $\Upsilon_V^{\rm dyn}$ Relations}
\label{sec:md_v_ml}
A number of interesting features shown by Figure~\ref{fig:mass_mld_size} regarding the mass, size and mass-to-light ratios of our sample GCs are of note.~We see a clear bifurcation in the $\Upsilon_V^{\rm dyn}$-${\cal{M}}_{\rm dyn}$ relations at ${\cal{M}}_{\rm dyn}\ga10^6\,M_\odot$, with two well defined sequences of GCs showing distinct positive slopes.~GCs below ${\cal{M}}_{\rm dyn}\!\approx\!10^6\,M_\odot$ do not seem to follow either of these two relations, appearing to have mass-to-light ratios in the range of $\sim\!1\!-\!5$ with no particularly well defined correlation.~In general, the circular points in Figures~\ref{fig:mass_mld_size} and \ref{fig:mag_mld_size} indicate a smooth transition from GCs with masses ${\cal{M}}_{\rm dyn}\!\approx\!10^5-10^6\,M_\odot$ to those with $\ga\!10^6\,M_\odot$ that follow the two sequences.

To compare our measurements with similar data of Local Group GCs we overplot measurements from \cite{mcl05} for Milky Way (MW), Large Magellanic Cloud (LMC), and Small Magellanic Cloud (SMC) GCs, along with data taken from \cite{str11} for M31 GCs.~It is important to point out here that while the \citeauthor{str11} measurements are based on direct kinematical measurements, the same cannot be said for the \cite{mcl05} data since they are based $\sigma_0$ estimates extrapolated from modelled light profiles.~Thus, while we use the \citeauthor{mcl05} data as a large, homogeneous comparison dataset, a true comparison cannot be made until ${\cal{M}}_{\rm dyn}$ measurements can be made for a large sample of MW GCs based directly on stellar kinematics. With that said, both samples align well with the bulk of our \cena\ GC sample but extend to significantly lower masses and fainter luminosities.~From the comparison with the majority of the Local Group GC sample, we conclude that the \cena\ GC sub-sample with ${\cal{M}}_{\rm dyn}\!\la\!10^6\,M_\odot$ and $\Upsilon_V^{\rm dyn}\!\la\!5\,M_\odot L_\odot^{-1}$ can be regarded as ``classical'' GCs similar to those found in the Local Group that follow the well-known GC ``fundamental plane'' relations  \citep{djo95, mcl00} where non-core-collapse GCs in the MW show almost constant core mass-to-light ratios of $\Upsilon_{V,0}=1.45\,M_\odot L_\odot^{-1}$.

\begin{figure}
\centering
\includegraphics[width=9.5cm, bb=60 0 1000 832]{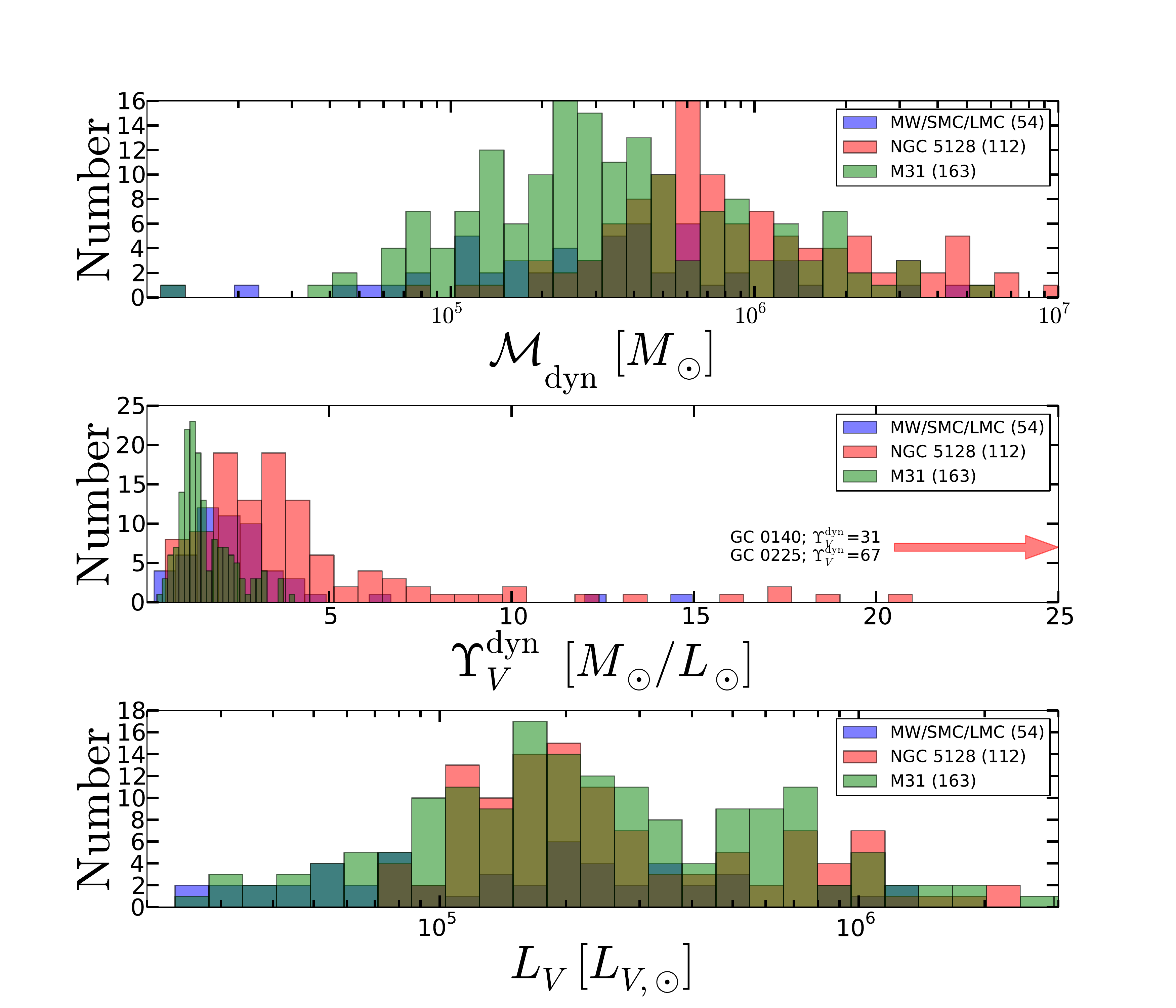}
\caption{Comparison between the ${\cal{M}}_{\rm dyn}$ {\it (top panel)}, $\Upsilon_V^{\rm dyn}$ {\it (middle panel)}, and $L_V$ {\it (bottom panel)} distributions for the \cena\ objects of the present work {\it (red histograms)}, and Local Group GCs (MW/SMC/LMC; {\it (blue histograms)}, and M31; ({\it green histograms})). Opacity has been lowered to show where the distributions overlap. While the MW/SMC/LMC sample is too small to make strong comparisons to the other two distributions, it is clear that the \cena\ sample shows higher median ${\cal{M}}_{\rm dyn}$ and $\Upsilon_V^{\rm dyn}$ than the M31 sample, despite probing a similar range in $L_V$. The total sample sizes are indicated in parentheses in the legends.}
\label{fig:ml_hist}
\end{figure}

It is difficult to establish whether the fundamental plane relations strictly hold for the ``classical'' \cena\ sub-sample as the core surface brightness values of these clusters are not accessible, even with {\it HST} imaging.~Even so, it can be seen in the middle panel of Figure~\ref{fig:ml_hist}, which compares the mass/light properties of our \cena\ sample with the Local Group GCs that the three distributions show a very similar rise up to $\Upsilon_V^{\rm dyn}\!\approx\!2\,M_\odot L_\odot^{-1}$; above which the \cena\ sample dominates.~With a MW/SMC/LMC sample that under-represents the peak of the GCLF at $L_{V,\odot}\!\approx\!2\cdot10^5\,L_\odot$ (see bottom panel of Figure \ref{fig:ml_hist}), it is beyond the scope of this work to verify whether a larger sample of MW/SMC/LMC GCs would ``fill in'' the distribution shown by our \cena\ sample.~The M31 sample, on the other hand, tends toward lower ${\cal{M}}_{\rm dyn}$, despite sampling a similar luminosity range as our \cena\ data.~We thus, for now, consider the \cena\ GC sub-sample shown by the circular points in Figures~\ref{fig:mass_mld_size} and \ref{fig:mag_mld_size} to simply represent the GCs that populate the Local Group.

More generally, we note that most \cena\ GCs have $\Upsilon_V^{\rm dyn}$ greater than the median value of MW/SMC/LMC GCs ($\Upsilon_{\rm MW}^{\rm dyn}\!\approx\!2.2\,M_\odot L_\odot^{-1}$) and significantly higher values than for M31 GCs ($\Upsilon_{\rm M31}^{\rm dyn}\!\approx\!1.37\,M_\odot L_\odot^{-1}$), although the latter may be due to insufficient aperture corrections for extended GCs \citep[see discussion in][]{str11}.~With a median $\Upsilon_V^{\rm dyn}$ of 3.44$\,M_\odot L_\odot^{-1}$, it is tempting to suggest a fundamental difference between Local Group GCs and those in \cena.~We do not strictly support this notion, as a more straight-forward interpretation is that we are simply sampling part of the GCMF that is inaccessible in the Local Group due to a dearth of known GCs above $10^6\,M_\odot$.~Indeed, the top panel of Figure~\ref{fig:ml_hist} demonstrates that the ${\cal{M}}_{\rm dyn}$ distributions of the Local Group and \cena\ GC samples are very dissimilar; the highest values are reached by \cena\ GCs, followed by M31 and the MW/SMC/LMC.~Above ${\cal{M}}_{\rm dyn}\!\approx\!5\cdot10^5\,M_\odot$, the \cena\ sample is well represented up to $\sim\!10^7\,M_\odot$, while Local Group GCs are more populous toward the lower tail of the GCMF. Altogether, Figure \ref{fig:ml_hist} suggests that the \cena/Local Group GCSs may have fundamentally different GCMFs, given the similarly sampled GCLF (bottom panel), something that can be tested when similar \cena\ data probing fainter magnitudes becomes available.

Having addressed the main, ``classical'' body of GCs in Figure \ref{fig:mass_mld_size}, we now turn to the two distinct high-${\cal{M}}_{\rm dyn}$ sequences. In the following we refer to GCs with $\Upsilon_V^{\rm dyn}\ga6$ and luminosities fainter than $M_V=-8.5$ mag (see Figure~\ref{fig:mag_mld_size}) as members of the ``dark star cluster'' (\hdyn) sequence (red dashed line in Figure~\ref{fig:mass_mld_size}) due to their potential connection to DSCs predicted by theory \citep[see \S\ref{sec:imbh};][]{ban11}. Those with ${\cal{M}}_{\rm dyn}\ga2\cdot10^6\,M_\odot$ and $\Upsilon_V^{\rm dyn}\la10$ that follow the shallower $\Upsilon_V^{\rm dyn}$-${\cal{M}}_{\rm dyn}$ relation (blue dashed line in Figure~\ref{fig:mass_mld_size}), we refer to as members of the ``dwarf-globular transition object'' (DGTO) branch, since these objects encroach upon the structural parameter space of the DGTOs reported by \cite{has05}.

Interestingly, the two objects omitted by these criteria, GC\,0217 and GC\,0239, lie intermediate between these two sequences and so we refer to them as ``intermediate-$\Upsilon_V^{\rm dyn}$" star clusters (see Figure~\ref{fig:mass_mld_size}).~Their properties may indicate either an evolutionary connection to one of the sequences, or perhaps represent a separate population that is simply not well sampled by our data.~In terms of structural parameters, the mean $r_h$ of the \hdyn\ sequence ($\langle r_h \rangle_{\rm high}\!=\!5.10$\,pc) is marginally smaller than that of the \ldyn\ branch ($\langle r_h \rangle_{\rm low}\!=\!6.67$\,pc), while also smaller in the median (4.01 and 5.59\,pc, respectively).~Meanwhile, the mean galactocentric radii of the two populations are $\langle R_{\rm gc}\rangle_{\rm high}\!=\!6.97\arcmin$ and $\langle R_{\rm gc}\rangle_{\rm low}\!=\!9.11\arcmin$.~{\sc Welch} 2-sample tests yield that the mean differences in $r_h$ and $R_{\rm gc}$ are not statistically significant, with p-values of 0.27 and 0.15, respectively.

\subsection{Properties of the $\Upsilon_V^{\rm dyn}$-${\cal{M}}_{\rm dyn}$ Sequences}
\label{sec:origins}
To probe the properties of the two $\Upsilon_V^{\rm dyn}$-${\cal{M}}_{\rm dyn}$ sequences, we fit empirical power-law relations of the form,
\begin{equation}
 \Upsilon_V^{\rm dyn}\propto{\cal{M}}_{\rm dyn}^\alpha
 \end{equation}
to approximate the data.~In a similar analysis of dispersion supported CSSs, including a subsample of the GCs considered here, \cite{tay10} found a value of $\alpha\!=\!0.24\!\pm\!0.02$ to fit their data, connecting ``classical'' GCs to more massive systems like UCDs and dwarf elliptical galaxies.~This relation is shown in Figure~\ref{fig:mass_mld_size} by the thin dashed black line and is too shallow to fit either of the $\Upsilon_V^{\rm dyn}$-${\cal{M}}_{\rm dyn}$ sequences of the present GC sample.~To better represent the data, we instead make an effort to fit power-laws to the sequences individually and find that each is well approximated by distinct, tight relations as described in the following.

The dashed blue line in Figure~\ref{fig:mass_mld_size} shows an approximation to the \ldyn\ sequence with a power-law slope of $\alpha\!=\!0.33\pm0.04$ (pentagons in all relevant Figures).~This relation fits the data quite well from the high mass GCs down to $\sim\!10^5\,M_\odot$, which represents a value more typical of Local Group GCs. Meanwhile, the \hdyn\ sequence (square points in all relevant Figures), shown by the red dashed line in Figure~\ref{fig:mass_mld_size} with a steeper slope ($\alpha=0.79\pm0.04$), seems to be created by a fundamentally different collection of objects.~Interestingly, we find two LMC GCs (NGC\,2257 and NGC\,1754, see Figure~\ref{fig:mass_mld_size}) that appear to align well with the \hdyn\ sequence.~While no strong statements can be made about only two objects, their exclusive presence around a currently interacting satellite of the MW marks an interesting starting point to investigate any connection to the \hdyn\ sequence.

$\Upsilon_V^{\rm dyn}$ is plotted as a function of $M_V$ in Figure~\ref{fig:mag_mld_size}, which shows that the \ldyn\ sequence is composed exclusively of the brightest GCs of the sample.~Thus, the \ldyn\ sequence may simply be explained by these GCs representing the high-luminosity tail of the GCLF.~On the other hand, GCs on the \hdyn\ sequence are fainter than \ldyn\ GCs by $\Delta M_V\!\approx\!1$ mag, making them similar in luminosity to the average GCLF turn-over magnitude found in many GC systems.~Furthermore, it can be seen that the range of $r_h$ for the objects on the \hdyn\ sequence is not dramatically different from that of the ``classical'' or \ldyn\ GCs.~Collectively, the similarity shown by these \hdyn\ objects in luminosity and size to other GCs in many GCSs likely explains why they have not been identified before in other galaxies, as they are only remarkable in their stellar dynamics properties.~Regardless, as this is the first time a clear distinction between two such groups of CSSs has been made, this naturally leads to the question of whether these objects should be called GCs at all.

\subsection{Possible Origins of the $\Upsilon_V^{\rm dyn}$-${\cal{M}}_{\rm dyn}$ Sequences}
Having shown artificial biases to be unlikely drivers of our results (see Appendix), the following discusses several astrophysical mechanisms that may generate our observations.

If the \ldyn\ sequence is made up of the brightest ``classical'' GCs, the $\Upsilon_V^{\rm dyn}\!\simeq\!5\!-\!9$ values shown by some are still perplexing in that GCs with $\Upsilon_V^{\rm dyn}\!\gtrsim\!5$ require additional explanations beyond being the extension of the ``classical" GCLF.~Effects that can mimic higher than usual $\Upsilon_V^{\rm dyn}$ include non-equilibrium dynamical processes (e.g.~rotation, pre-relaxation, young stellar populations, tidal disruption) or an exotic IMF.~We find several \ldyn\ sequence GCs with ellipticities, $\epsilon\ga0.25$, indicative of a non-equilibrium dynamical state, such as rotation. Other GCs show young ($\la\!8$ Gyr) ages \citep{woo10b}, and thus may not be fully relaxed.~Additionally, for four objects on this \ldyn\ sequence (GC\,0041, GC\,0330, GC\,0365, and GC\,0378) \cite{har02a} found evidence for extra-tidal light contributing to their surface-brightness profiles in excess of their King model fits.~For the remainder, the possibility of a particularly bottom-heavy IMF \citep[e.g.][]{dab08,mie08b} could explain their elevated $\Upsilon_V^{\rm dyn}$ estimates.

\subsubsection{Globular Cluster Rotation}
\label{sec:gc_rotation}
To probe how non-equilibrium states could explain the elevated $\Upsilon_V^{\rm dyn}$, we investigate the possible impact of rotation on our mass estimates.~Treating the observed $\Upsilon_V^{\rm dyn}$ as being strictly the result of the GCs exhibiting a mass excess at a given luminosity with respect to the median $\Upsilon_{\rm MW}^{\rm dyn}\!=\!2.2$, we calculate for each of our sample GCs the amount of ``extra'' mass within $r_h$, or $\Delta\!{\cal{M}}_{1/2}=\!\left(\Upsilon_V^{\rm dyn}\!-\!\Upsilon_{\rm MW}^{\rm dyn}\right)\cdot\!\frac{L_V}{2}=\!\left(\Upsilon_V^{\rm dyn}\!-\!2.2\right)\cdot\!\frac{L_V}{2}$.~This mass component then needs to be accounted for by the effects discussed above to explain the elevated $\Upsilon_V^{\rm dyn}$ values.

\begin{figure*}[t]
\centering
\includegraphics[width=8.9cm, bb=0 0 850 850]{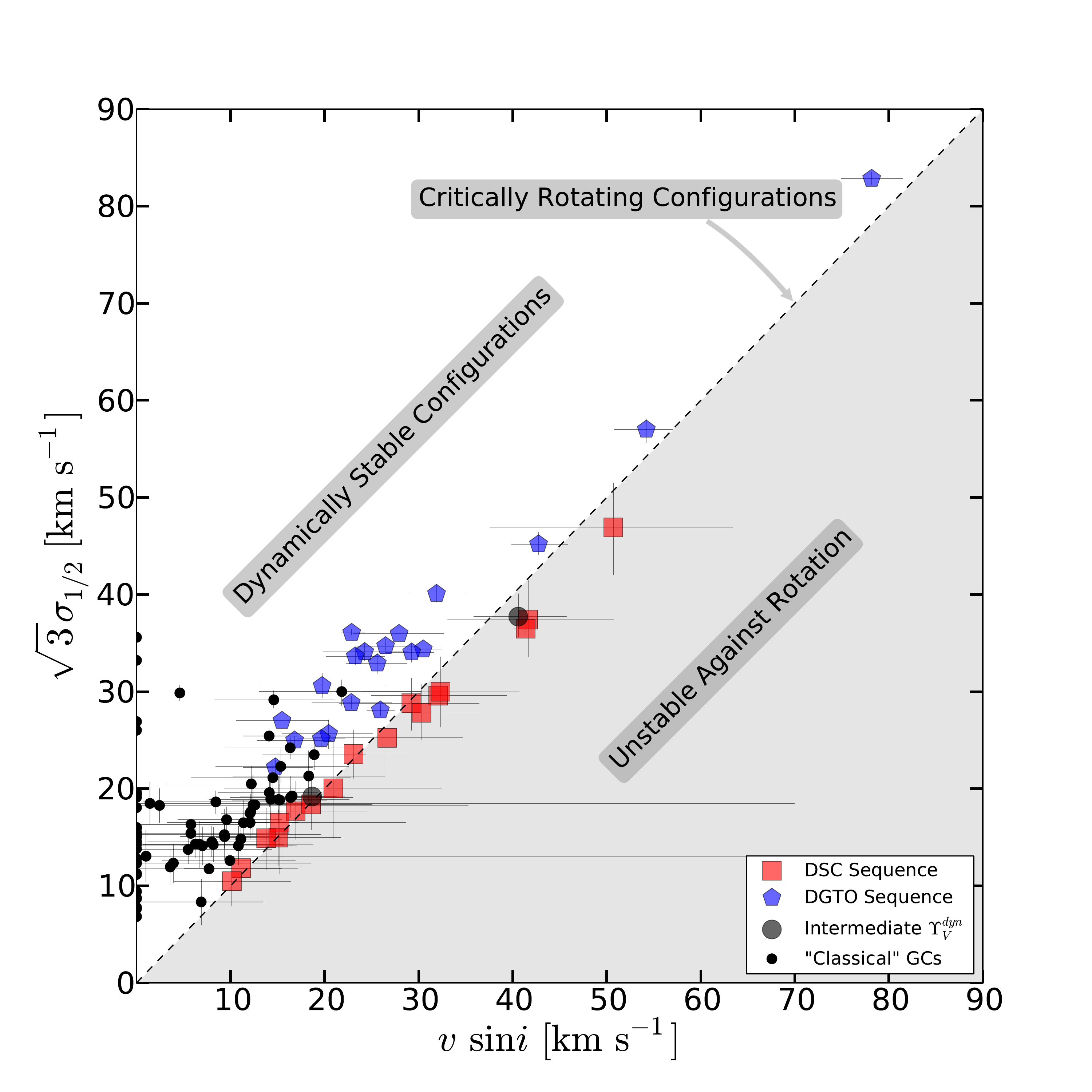}
\includegraphics[width=8.9cm, bb=0 0 850 850]{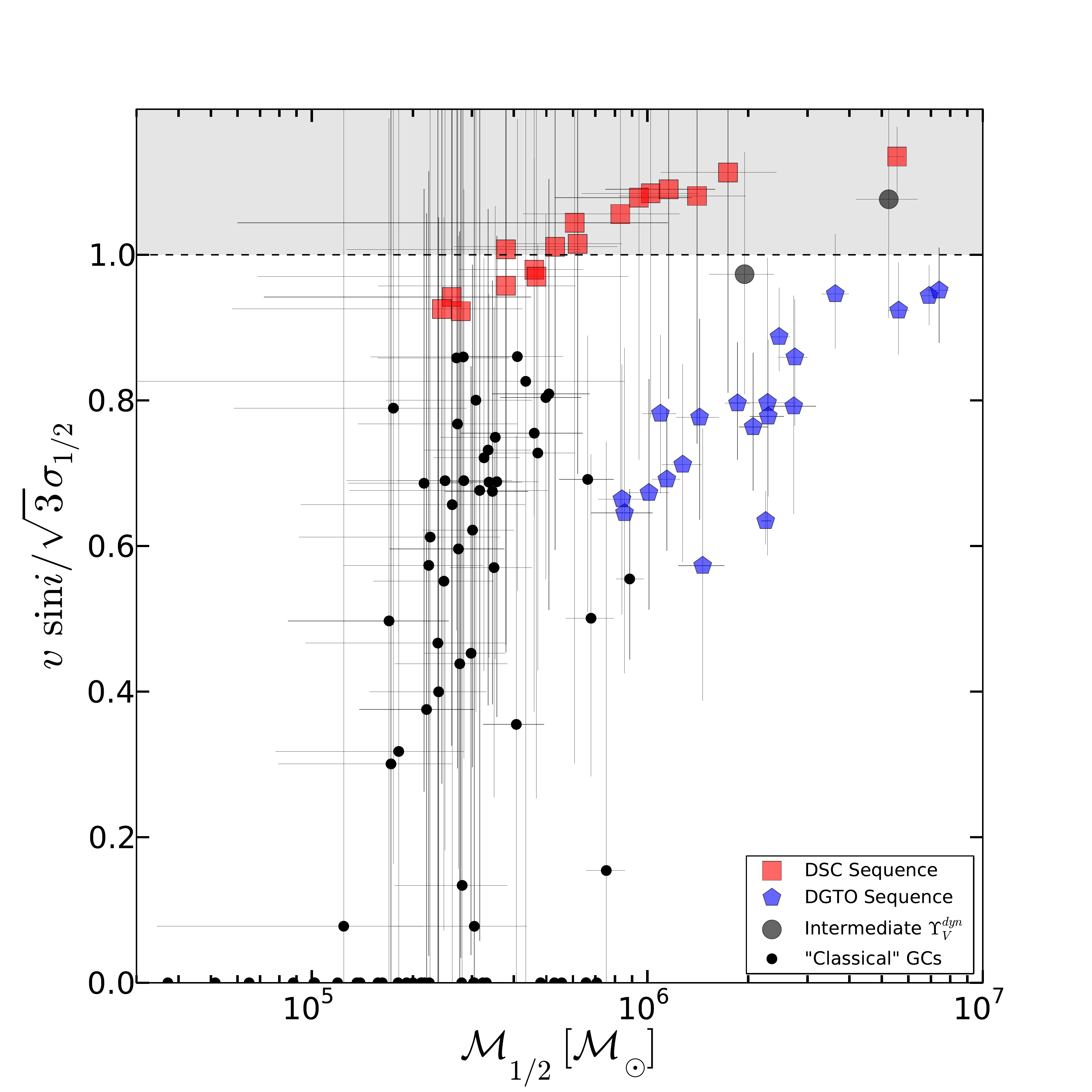}
\caption{({\it Left panel}): Average measured orbital velocities of cluster stars shown as a function of the minimum circular velocities necessary for rotation alone to account for the observed mass-to-light ratios.~The dashed line represents the critical velocity (assuming $\sin\,i\!=\!1$, or the maximum observable velocities) at which {\it a)}\,rotation alone is able to explain the elevated $\Upsilon_V^{\rm dyn}$ of an object, or/and {\it b)}\,an object of a given mass becomes unstable against rotation, consistent with a dynamical non-equilibrium state. Symbol shapes are as in Figures \ref{fig:mass_mld_size} and \ref{fig:mag_mld_size}, but with the ``classical'' GCs shown as points and the two intermediate-$\Upsilon_V^{\rm dyn}$ GCs (GC\,0217 and GC\,0239, see Section~\ref{sec:md_v_ml}) shown as gray symbols. ({\it Right panel}): $v\sin i/\sqrt{3}\sigma_{1/2}$ as a function of the dynamical half-mass, ${\cal{M}}_{1/2}$, for the same objects.~There are correlations for the objects on the \hdyn\ and \ldyn\  sequences, with a weaker correlation shown by the ``classical'' GCs with $v\sin i/\sqrt{3}\sigma_{1/2}\!\ga\!0.2$, indicating that GCs become more rotationally supported at higher masses.}
\label{fig:rotation}
\end{figure*} 

Making the naive assumption that $\Delta\!{\cal{M}}_{1/2}$ is entirely due to rotation, then in the general situation where the rotation axis is aligned at any angle $i$ with respect to the observational plane, stars at $r_h$ would require circular velocities of at least,
\begin{equation}
v\sin i\!\simeq\!\left(\frac{\Delta\!{\cal{M}}_{1/2}G}{r_h}\right)^{1/2}.
\end{equation}

To investigate how rotation could explain the $\Upsilon_V^{\rm dyn}$ estimates, the left panel of Figure~\ref{fig:rotation} shows a comparison between $v\sin i$ and the average stellar velocity in GCs computed from the random stellar motions via $\sigma_{1/2}\sqrt{3}$.~The symbols are the same as in Figures~\ref{fig:mass_mld_size} and \ref{fig:mag_mld_size}, but with ``classical'' GCs simply shown as dots.~The dashed line in Figure~\ref{fig:rotation} indicates the boundary at which rotation is equal to the random stellar motion component required to be consistent with $\Delta\!{\cal{M}}_{1/2}$.~It must be acknowledged that the errors bars shown on Figure \ref{fig:rotation} make it impossible to definitively discuss the dynamical configurations of our sample. On the other hand, the lack of co-mingling between the different groups suggests that the features seen are probably not solely due to systematic errors, thus they can still be used to make general statements about the populations as wholes, and we proceed with that in mind.

A GC that has a rotation to random motion component ratio, $v\sin i/\sigma_{1/2}\sqrt{3}\!>\!1$, requires circular velocity speeds that would destabilize the system if only rotation is to explain its $\Delta\!{\cal{M}}_{1/2}$.~Most \hdyn\ objects fall on or below this unity relation. These clusters are generally consistent with non-equilibrium dynamical configurations, and may require at least one other effect to explain their high $\Upsilon_V^{\rm dyn}$ values, e.g.~dark gravitating components.~On the other hand, a GC with $v\sin i/\sigma_{1/2}\sqrt{3}\!<\!1$ can have net angular momentum that can provide a stable configuration against rotational breakup.~In this case, GC rotation alone can account for their $\Upsilon_V^{\rm dyn}$.~Significant error bars notwithstanding, all of the ``classical'' GCs along with those on the \ldyn\ sequence are exclusively consistent with $v\sin i/\sigma_{1/2}\sqrt{3}\!<\!1$.~Thus, their elevated $\Upsilon_V^{\rm dyn}$ could be explained by rotation without the need to invoke additional components, bolstering the interpretation that they represent the high-luminosity tail of \cena's GCLF.

As a corollary of the previous exercise, we investigate in the right panel of Figure~\ref{fig:rotation} whether the ratio $v\sin i/\sigma_{1/2}\sqrt{3}$ correlates with ${\cal{M}}_{1/2}$ (see Section~\ref{sec:masses}).~We find that the \hdyn\ and \ldyn\ GC samples appear to exhibit correlations in this parameter space, hinting at rotational support that increases with GC mass.~This result concurs with the recent findings of \cite{kac14} and \cite{fab14}, who measure small but significant rotational speeds in MW GCs. However, given that $\sin i$ is likely to be randomly distributed in the range [0..1] (hence, $\langle\sin i\rangle\!=\!2/\pi$), these correlations, if real and due to rotational support alone, should be much noisier than what we see in the right panel of Figure~\ref{fig:rotation}, and are therefore probably driven by other effects than rotation alone.


\subsubsection{Central Massive Black Holes}
\label{sec:imbh}
A potential source of artificially enhanced ${\cal{M}}_{\rm dyn}$ values are the effects of central intermediate-mass black holes \citep[IMBHs; e.g.][]{saf10, mie13, lei14} of lesser mass, but otherwise not unlike that found recently in a UCD \citep{set14}.~To estimate the influence of a putative central compact object we compute expected IMBH masses by using our $\sigma_0$ estimates with the BH mass vs.~velocity dispersion relation, ${\cal{M}}_{\rm BH}\!-\!\sigma$, for CSSs, which is offset from that of pressure supported galactic systems \citep[][their Figure 6]{mie13}.~From the ${\cal{M}}_{\rm BH}\!-\!\sigma$ relation, assuming that it scales to lower-mass stellar systems, we obtain BH masses in the following ranges $6.6\cdot10^4\la{\cal{M}}_{\rm BH}/M_\odot\la5.0\cdot10^7$ for the \ldyn\ GCs, $8.5\cdot10^2\la{\cal{M}}_{\rm BH}/M_\odot\la2.2\cdot10^6$ for the \hdyn\ objects, and ${\cal{M}}_{\rm BH}\la7.6\cdot10^5\,M_\odot$ for the combined ``classical'' and intermediate-$\Upsilon_V^{\rm dyn}$ sample.

\begin{figure}[t]
 \includegraphics[width=8.9cm, bb=0 0 850 850]{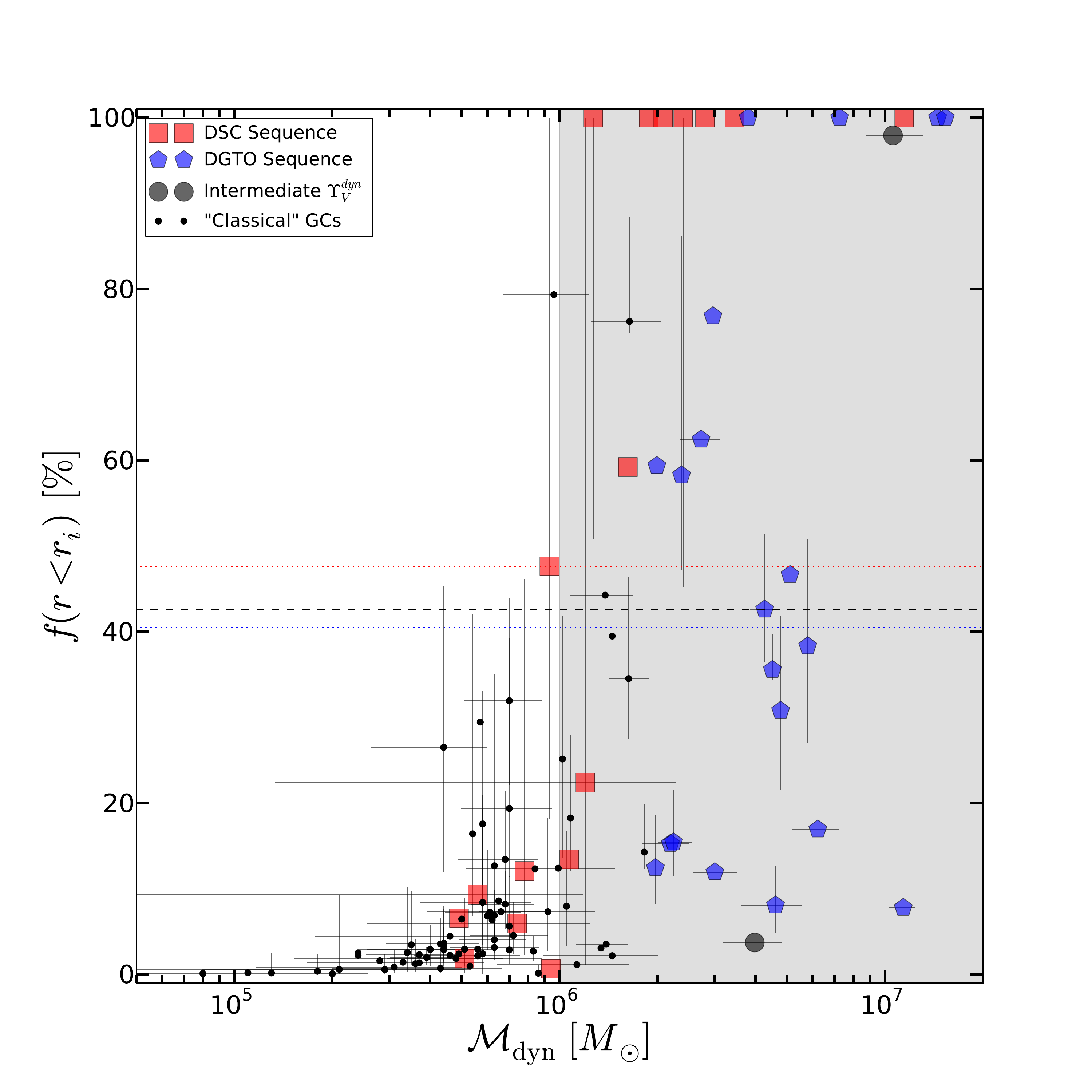}
 \caption{Fraction of the stellar mass within the sphere of influence of a putative central IMBH, as a function of ${\cal{M}}_{\rm dyn}$ for our entire sample. The black dashed line shows the median of the combined \ldyn\ and \hdyn\ sequence GCs, while the red- and blue-dotted lines show the medians for the individual sub-samples.}
 \label{fig:imbh}
\end{figure}

For each GC we integrate the central stellar light profile using our numerical models (see \S \ref{sec:apcorr}) until the radius of the sphere encompasses a stellar mass corresponding to $2{\cal{M}}_{\rm BH}$, assuming the median $\Upsilon_{\rm MW}^{\rm dyn}\!=\!2.2$.~This radius defines the IMBH sphere of influence \citep[$r_i$;][]{mer04}.~We then compute the fraction of the stars within this sphere with respect to the modeled population falling within our apertures, $f(r\!<\!r_i)$.~In Figure~\ref{fig:imbh}, we plot this fraction as a function of ${\cal{M}}_{\rm dyn}$ for our entire sample.

While most of the ``classical'' and intermediate-$\Upsilon_V^{\rm dyn}$ GCs show little dynamical influence by potential IMBHs, there is an upturn in $f(r\!<\!r_i)$ for GCs with ${\cal M}_{\rm dyn}\!\ga\!10^6\,M_\odot$.~The high-mass sub-sample is mostly made up of the \ldyn\ and \hdyn\ objects.~A {\sc Mood} test \citep{moo50} for equal medians (see horizontal lines in Figure~\ref{fig:imbh}) provides no evidence for a difference in the \ldyn\ and \hdyn\ sequence $f(r\!<\!r_i)$ (p-value=1.00).~Specifically, Figure~\ref{fig:imbh} shows a bimodal distribution in $f(r\!<\!r_i)$ for the \hdyn\ sequence objects, as 7/17 and 8/17 fall either at the $\sim\!100\%$ or $\la\!20\%$ levels, respectively.~Two of these objects show intermediate $f(r\!<\!r_i)\!\simeq\!50\!-\!60\%$.~For the \ldyn\ sequence GCs there is a smoother transition from those with little to no influence by a putative IMBH (7/20) to those which would have the majority of their stars dynamically dominated by such an object (8/20).~Taken together, the presence of central IMBHs could in principle provide a plausible explanation for the observed dynamics of many objects on both sequences.


We point out that some GCs in the grey shaded region in Figure~\ref{fig:imbh} are identified as X-ray sources, almost all of which are classified as Low-Mass X-ray Binary (LMXB) hosts from {\it Chandra} observations \citep{liu11}.~2/17 \hdyn\ objects are X-ray sources, compared to 8/20 \ldyn\ GCs.~This is in line with the following argument: given the fainter nature of the \hdyn\ objects, they must have fewer stars compared to \ldyn\ clusters at a given ${\cal{M}}_{\rm dyn}$.~With fewer stars providing stellar winds/mass-loss, one would generally expect accretion onto an IMBH/LMXB to be less likely compared to \ldyn\ sources.

Given the apparently enigmatic properties of the faint \hdyn\ subsample, it is important to note that BHs dominating their dynamics might alter our basic assumption of the canonically accepted \cite{kin66} stellar density profile.~Spatially resolved $\sigma$ profiles would test the putative central BH sphere of influence and thus the validity of our assumption. Compared to a uniformly distributed mass component, a central IMBH can mimic a dynamical mass as much as $4-5\times$ higher than the mass of the BH itself \citep{mie13}. Scaling $\Delta\!{\cal{M}}_{1/2}$ in this sub-sample down by four then suggests that BHs of masses $3.9\cdot10^4\la{\cal{M}}_{\rm BH}/M_\odot\la1.3\cdot10^6$ could plausibly provide the $\sigma_{\rm ppxf}$ that we observe, as well as the elevated $\Upsilon_V^{\rm dyn}$ values.~In any case, the lack of a strong correlation in the $f(r\!<\!r_i)$ vs.~${\cal{M}}_{\rm dyn}$ plane calls into question whether IMBHs would be the only driver of the two sequences.

Despite the plausibility of single massive central BHs explaining some of our results, the potential connection to DSCs predicted recently by \cite{ban11} needs to be considered.~In this scenario, neutron star and BH remnants of massive stars gather as a very concentrated central dark sub-cluster.~Passages through a strong tidal field act to strip luminous matter from the outskirts, resulting in very high mass-to-light ratios.~To be observable, the stellar stripping process must act on timescales shorter than the self-depletion of dark remnants via encounter-driven mechanisms (e.g. three-body interactions). \cite{ban11} predict lifetimes of such objects with stellar masses $\la\!10^5\,M_\odot$ orbiting within 5\,kpc of a MW-like potential to be generally less than 1\,Gyr, calling into question the likelihood of observing such a current population. With that said, we note that given our DSC stellar masses of ${\cal{M}}_*\simeq1-5\cdot10^5\,M_\odot$, and $R_{\rm gc}\ga5\,{\rm kpc}$, combined with the predicted correlation between lifetime, and both ${\cal{M}}_*$ and $R_{\rm gc}$, these clusters could plausibly have DSC phases lasting on Gyr timescales. Detailed future modeling will be critical in determining the plausible parameter space necessary for the existence of these objects given the unique history of \cena.

\subsubsection{Dark Matter Halos}
\label{sec:dm}
If central IMBHs are not driving the dynamics of the \hdyn\ sequence, then explaining objects with $\Upsilon_V^{\rm dyn}\!\ga\!10$ becomes very difficult without requiring a significant amount of dark matter (DM). It is generally accepted that ``classical'' GCs are devoid of non-baryonic matter, but this cannot be entirely ruled out \citep[e.g.][]{moo96, con11, sol12, iba13} and may in fact be expected from theoretical considerations \citep[e.g.][]{pee84, sai06}. Having considered and ruled out inflated $\Upsilon_V^{\rm dyn}$ values due to myriad observational/instrumental effects (see Appendix), and/or severely out-of-equilibrium dynamical states (see Section~\ref{sec:gc_rotation}) due to, e.g.,~rotation, we now consider the implications of the \hdyn\ sequence being due to the onset of DM domination in low-mass systems (${\cal M}\!\ga\!10^5\,M_\odot$). 

\begin{figure}
 \includegraphics[width=\linewidth]{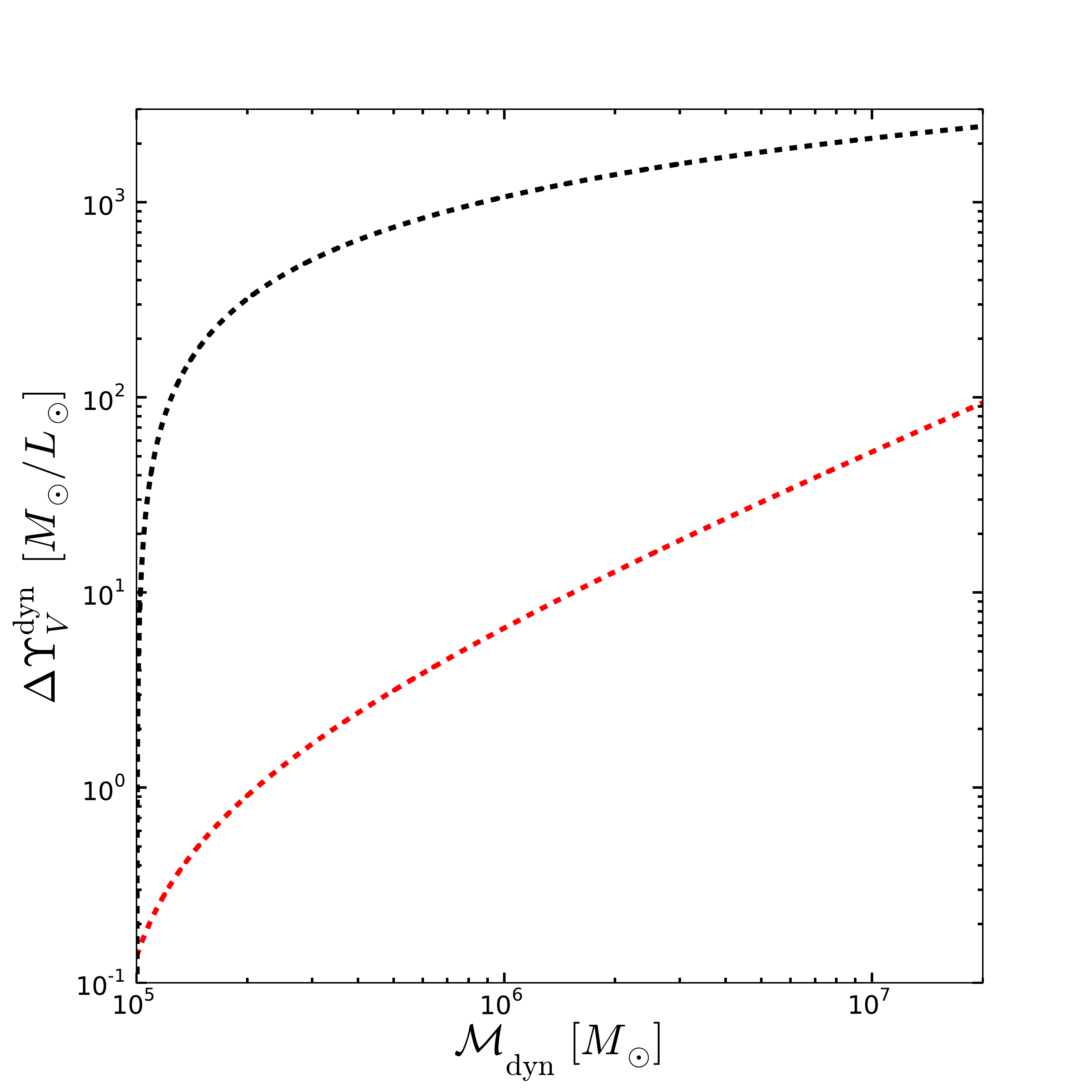}
 \caption{Difference in $\Upsilon_V^{\rm dyn}$ between the red- and blue-dashed lines in Figure~\ref{fig:mass_mld_size} (red relation), and the observational limit of our data (black relation) as a function of ${\cal{M}}_{\rm dyn}$. The dominance of the dark gravitating mass component exponentially increases above a truncation mass of a few times $10^5\,M_\odot$.}
 \label{fig:xmass_frac}
\end{figure}

The red-dashed line in Figure~\ref{fig:xmass_frac} shows the difference in $\Upsilon_V^{\rm dyn}$ between the \ldyn\ and \hdyn\ power-law fits as a function of $\cal{M}_{\rm dyn}$. In other words, this relation shows the amount of DM required to explain the exponentially increasing mass excess compared to ``classical'' GCs.~If dark matter is behind the \hdyn\ objects, then the sharp truncation at ${\cal M}_{\rm dyn}\!\approx\!10^5\,M_\odot$ hints at a population of increasingly DM dominated structures in the immediate vicinity of \cena\ with masses as low as a few times $10^5\,M_\odot$.

Recent modeling has shown that in a realistic galaxy cluster potential, as much as 80-90\% of a dwarf galaxy's DM halo may be stripped before baryonic losses become observable \citep{smi13a}. If the \hdyn\ objects originate from low-mass halos that have been stripped during their passage(s) through \cena's potential well, this then implies progenitors of $\cal{M}_{\rm dyn}$ (and thus $\Upsilon_V^{\rm dyn}$) $\sim\!10\times$ higher than we observe.~This scenario would suggest that at one point in time during the assembly history of \cena\ there may have been a population of compact, low-luminosity baryonic structures inside the virial radius of \cena\ embedded within extraordinarily compact dark matter halos of $10^7\!\la\!\cal{M}_{\rm dyn}/M_\odot\!\la\!$ $10^8$ with $100\!\la\!\Upsilon_V^{\rm dyn}/M_\odot L_\odot^{-1}\!\la700$ \citep[e.g.][]{ric03}.

If the above is true, then the truncation of the red-dashed line in Figure~\ref{fig:xmass_frac} at $\sim\!10^5\,\cal{M}_{\rm dyn}$ may mark the limit below which primordial DM halos have not survived accretion events onto larger galaxy structures.~This result is consistent with the picture recently put forward in \cite{puz14} where the sizes of outer halo star clusters are truncated by the abundance of small DM halos \citep[see also][]{car09, car11, car13}. In this scenario, CSSs of larger masses suffer higher levels of dynamical friction than low-mass clusters while moving through the potential well of the host galaxy and sink closer in to the central body \citep[e.g.][]{lotz01}.~While sinking, they suffer extra harassing encounters in the more crowded core region, which act to further truncate their sizes \citep[e.g.][]{webb13}.~The above may be the explanation for why the \hdyn\ objects have sizes typical of Milky Way GCs (see Figures~\ref{fig:mass_mld_size} and \ref{fig:mag_mld_size}, and also Figure~16 in \citealt{puz14}).

\subsection{Potential Progenitors of the DSC Sequence GCs}
Figure~\ref{fig:mag_mld_size} shows that with magnitudes of $-8.5\!\la\!M_V\la-7.4$ mag, the \hdyn\ objects are among the intrinsically faintest star clusters of our sample.~This range scratches the peak of the GCLF and begins to infringe upon the realm of ultra-faint dwarf galaxies (UFDs) in the Local Group, which are thought to be the extension of the dwarf spheroidal (dSph) galaxy population to lower luminosities \citep[$M_V\!\ga\!-8$ mag; e.g.,][]{wil05,zuc06,bel07,zuc07,mcc09,bro12}.~Despite smaller sizes than known Local Group dSphs, the combination of high $\Upsilon_V^{\rm dyn}$ and low luminosities shown may suggest an evolutionary link to a putative population of dSph- and/or UFD-like, dark-matter dominated dwarf galaxies.~If so, they must have undergone an extremely compact and therefore efficient early star-formation burst before being tidally stripped of most of their dark-matter halos during subsequent interactions with \cena\ \citep[e.g.][]{smi13a, gar14, saw14}.

The fact that we do not observe current tidal features indicative of stripping might be due to the surface brightness limits of available data, but higher (i.e.~non-equilibrium) LOSVDs might still be consistent with theoretical predictions.~For instance, \cite{smi13b} model a dwarf galaxy similar to the UFD UMaII, which with $\sigma\!=\!6.7\, {\rm km}\,{\rm s}^{-1}$ \citep[implying $\Upsilon_V^{\rm dyn}\!\approx\!1000$;][]{sim07} is an obvious likely candidate for a \hdyn\ sequence object progenitor. By subjecting it to tidal forces in a Milky Way-like gravitational potential, these authors successfully reproduced many observed properties (e.g.~luminosity, central surface brightness, ellipticity, etc.) of UMaII, and found that the galaxy's $\sigma$ can be boosted on timescales of a few Gyr, in particular around the apocentre of the orbit.~Their modeling showed that ``$\sigma$ boosting'' can easily reproduce UMaII's $\sigma$, and can even reach extreme levels of $>\!20$ km s$^{-1}$ when the orbital trajectory is close to being perpendicular to the disk of the host galaxy.~The models therefore suggest that the extreme $\Upsilon_V^{\rm dyn}$ for the \hdyn\ objects could potentially be explained without requiring large amounts of DM.~While this result, assuming that ``$\sigma$ boosting'' scales to a gE galaxy like \cena, might be sufficient to explain a handful of the \hdyn\ clusters, we consider it highly unlikely that all such clusters can be accounted for, due to the required synchronization of their apocentre passages and current line-of-sight alignments.~Note that this mechanism would generally introduce scatter rather than producing the sequences observed in Figure~\ref{fig:mass_mld_size}.

Assuming that ``$\sigma$ boosting'' is unable to solely account for the \hdyn\ objects, then the small sizes compared to UFD-like galaxies call for careful skepticism if DM is the preferred explanation.~For example, characteristic central DM densities for dwarf galaxies, if canonical halo profiles are assumed, can range from $0.1\,M_\odot$ pc$^{-3}$ for preferred cored profiles, to as much as $60\,M_\odot$ pc$^{-3}$ for cuspy profiles \citep[e.g.][]{gil07,tol11}.~In this case, the required DM masses (${\cal{M}}_{\rm DM}$) of $1.6\cdot10^5<{\cal{M}}_{\rm DM}/M_\odot<5.4\cdot10^6$ and corresponding densities ($\rho_{\rm DM}$) of $40<\rho_{\rm DM}/M_\odot\, \mbox{pc}^{-3}<4.5\cdot10^4$ within the \hdyn\ cluster half-light radii are at least three orders of magnitude higher than those expected from a cuspy dwarf galaxy profile, or if cored, they would each need to be embedded in $\sim10^{14}-10^{15}\,M_\odot$ halos.

Irrespective of this problem, such high DM densities might actually give rise to central baryonic concentrations that enable the formation of IMBHs.~This might be realized by triggering an extremely dense central stellar environment, leading to a runaway collisional event that culminates with the formation of a central BH that might continue to grow through the accretion of binary star and higher-order multiplets.

Given all of these interpretations, it seems just as likely that a combination of central IMBHs and/or dark stellar remnants, cuspier-than-expected DM halos, and/or ``$\sigma$ boosting'' could be at work.~This conclusion may in fact be the most reasonable one, given that IMBHs would presumably be of dwarf galaxy origin given the difficulty in building up such a mass in a GC-like structure without just as remarkable early star-formation efficiency.~Then, the residual DM leftover after stripping, if cuspy, in combination with ``$\sigma$ boosting'', may be sufficient to produce these new objects.~This final interpretation would then require both less extreme BH and DM masses, while avoiding the orbital synchronization problem.~Regardless of the mechanism giving rise to them, the emergence of the \hdyn\ sequence objects is a very unexpected result and calls for detailed follow-up observations and high spatial resolution modeling.

\section{Summary \& Future Outlook}
\label{sec:conc}
New high-resolution spectra of compact stellar objects around the giant elliptical galaxy \cena\ (Centaurus A) were analyzed.~We combined these data with a re-analysis of 23 clusters from the literature and used a penalized pixel fitting technique to derive new radial velocities for 125 objects (3 first-time measurements), as well as line-of-sight velocity dispersions for 112 targets (89 first-time measurements).~Based on these estimates we derived dynamical mass and mass-to-light ratio estimates by combining the new kinematical information with structural parameters (mostly obtained from HST imaging) and photometric measurements from the literature.

We briefly summarize our results as follows:
\begin{itemize}[leftmargin=*]
	\item At intermediate GC masses ($10^5\!\la\!{\cal M}_{\rm dyn}/M_\odot\!\la\!10^6$) we find the expected population of ``classical'' GCs, with no anomalous kinematical results.~These GCs resemble those of the Local Group in every way, albeit being slightly brighter than average due to our sample selection bias.
	\item At the high-mass end (${\cal M}_{\rm dyn}\!\ga\!10^6\,M_\odot$), {\it we find at least two distinct star-cluster populations} in the $\Upsilon_V^{\rm dyn}$-${\cal{M}}_{\rm dyn}$ plane which are well approximated by power-laws of the form $\Upsilon_V^{\rm dyn}\!\propto\!{\cal M}_{\rm dyn}^{\alpha}$.
	\begin{itemize}
		\item The ``\ldyn\ sequence'' is comprised of objects with ${\cal M}_{\rm dyn}\ga2\cdot10^6\,M_\odot$ and $\Upsilon_V^{\rm dyn}\la10\,M_\odot L_\odot^{-1}$ and is well described by a power-law with a slope $\alpha=0.33\pm0.04$.
		\item The ``\hdyn\ sequence'' objects have ${\cal M}_{\rm dyn}$ similar to the \ldyn\ sequence clusters, but with a significantly steeper power-law slope,~$\alpha\!=\!0.79\!\pm\!0.04$.~Moreover, the faint magnitudes ($M_V\!\ga\!-8.5\,{\rm mag}$) lead to anomalously high ($\Upsilon_V^{\rm dyn}\!\geq\!6\,M_\odot L_\odot^{-1}$) mass-to-light ratios.~We point out that at least two LMC GCs (NGC\,1754 and NGC\,2257) also appear to follow this relation.
	\end{itemize}
\end{itemize}

Despite being among the brightest clusters of our sample, some \ldyn\ sequence objects show relatively high mass-to-light ratios in the range $5\la\Upsilon_V^{\rm dyn}/M_\odot L_\odot^{-1}\la10$, which require explanation if a non-baryonic mass component is to be avoided. We find that extreme rotation and/or dynamically out of equilibrium configurations can explain their kinematics as well as indications of higher levels of rotational support with increasing ${\cal M}_{\rm dyn}$. Plausible alternative and/or concurrent explanations also include particularly top- or bottom-heavy IMFs, or the dynamical influence of central IMBHs. Altogether we consider that these objects represent the very bright tail of the GCLF which is well represented around \cena, but poorly populated in the Local Group.

While the \ldyn\ sequence has a fairly pedestrian explanation, the $6\la\Upsilon_V^{\rm dyn}/M_\odot L_\odot^{-1}\la70$ values of the \hdyn\ sequence objects are much more difficult to reconcile with the \ldyn\ branch scenarios, save for a small subset.~We investigated in detail (see Appendix, for details) the potential impact of observational, low S/N, and/or instrumental effects on artificially inflating the \hdyn\ values, and found that astrophysical explanations are required.

For most of these objects the average stellar velocities ({\it vis-\`a-vis} the observed $\sigma_0$ values) do not appear to be high enough for extreme rotation to explain their dynamics, and if they were, then the clusters would be unstable against rotational break-up. Moreover, it is highly unlikely that all clusters would have their rotation axes aligned with the plane of the sky (assuming $\sin i\!\in\! U[0,1]$), as would be needed to reproduce and minimize the scatter in the observed $\Upsilon_V^{\rm dyn}$ vs.~${\cal M}_{\rm dyn}$ relation.~Combined with the difficulty in explaining a mechanism that would impart and maintain the necessary angular momentum for such large rotational velocities, we thus consider extreme rotation/significantly out-of-equilibrium dynamical configurations insufficient and unlikely to explain their properties.

We considered the plausibility of central IMBHs and a central accumulation of dark stellar remnants, consisting of stellar-mass BHs and neutron stars, to be driving the extreme dynamics of the \hdyn\ clusters.~If they exist, then putative central IMBHs and remnant population can plausibly influence enough of the stars in many of the \hdyn\ clusters to provide an explanation for their velocity dispersion measurements.~In fact, if our assumed canonical structural profiles were sufficiently perturbed by an IMBH's presence and/or stellar remnant population, then this explanation requires even less massive IMBHs to become plausible.~With that said, given that the IMBH + stellar remnant interpretation can only account for the dynamics of $\sim50\%$ of these objects, this is unlikely to explain the emergence of the \hdyn\ sequence.

If central IMBHs and stellar remnant populations are not the only cause of the elevated $\Upsilon_V^{\rm dyn}$, then the possibility of significant dark matter mass components must be considered, despite the wide acceptance that ``classical'' GCs are devoid of DM.~This result would have important implications for GC formation models and early structure formation, and indicate that not all extragalactic star clusters are genuine GCs.~More importantly, the presence of such amounts of DM in the \hdyn\ sequence objects would imply that they represent the lowest mass primordial dark matter halos that have survived accretion onto larger-scale structures to the present day. In other words, there may still exist a large reservoir of $10^5\,M_\odot$-scale dark matter halos surviving in relative isolation in the universe today, at least around relatively quiescent larger dark matter halos like \cena.~Moreover, if these objects are stripped of formerly more massive dark matter halos, presumably as former dwarf galaxies, this would imply the presence of a significant collection of objects with $10^7\!\la\!{\cal M}_{\rm dyn}/M_\odot\!\la\!10^8$, and $100\la\Upsilon_V^{\rm dyn}/M_\odot L_\odot^{-1}\la1000$ in the relatively recent past of \cena.

This interpretation is not without its serious problems.~For example, with the above properties, central ($\sim\!10$\,pc) DM masses/densities on the order of $\sim\!10^3\times$ larger than canonical DM halo profiles would be required; a scenario that cannot be reconciled with any current theoretical framework.~Given the improbability that such massive central BHs, exotic and ultra-concentrated DM halos, or extremely out-of-equilibrium dynamical configurations can individually explain the properties of the \hdyn\ sequence objects, it seems perhaps most likely that a mixed bag of such factors may be at play behind this result, although it is puzzling how a combination of these physical mechanisms would conspire to generate a relatively sharp $\Upsilon_V^{\rm dyn}$ vs.~${\cal M}_{\rm dyn}$ relation.

It remains to be seen if similar objects exist in the star cluster systems of other galaxies, but verification of these results will be difficult for more distant systems due to their intrinsic low luminosities.~Nonetheless, detailed chemical abundance studies of these objects will shed light on the origins (e.g.~primordial or not, simple or multi-generational stellar populations, etc.)~of both \ldyn\ and \hdyn\ clusters.~While the proximity of \cena\ provides the possibility to study the internal dynamics and stellar populations of its CSSs, the distance approaches the limits of what is currently feasible with today's instrumentation on 8-10m class telescopes.~Nonetheless, large-scale, complete studies of the chemo-dynamics of GCs in the Local Group, and of giant galaxies within $\sim5\,{\rm Mpc}$, will help reveal the true nature of this enigmatic new type of compact stellar systems.

\acknowledgments

We wish to thank Andres Jord\'an for providing us with $r_h$ measurements based on {\it Hubble Space Telescope} observations prior to publication, and Tim-Oliver Husser for his help with the latest version of the {\sc PHOENIX} library.~We thank Roberto Mu\~noz, Mia Bovill, Jincheng Yu, Simon \'{A}ngel, Rory Smith, Graeme Candlish, Steffen Mieske, and Pavel Kroupa for fruitful discussions and comments that served to improve the manuscript.~We also extend our gratitude to the anonymous referee for constructive criticisms that significantly improved this work.~This research was supported by FONDECYT Regular Project Grant (No.~1121005) and BASAL Center for Astrophysics and Associated Technologies (PFB-06).~M.A.T.~acknowledges the financial support through an excellence grant from the ``Vicerrector\'ia de Investigaci\'on" and the Institute of Astrophysics Graduate School Fund at Pontificia Universidad Cat\'olica de Chile and the European Southern Observatory Graduate Student Fellowship program.~M.G.~acknowledges financial support through Project ``Nucleo de Astronomia", Universidad Andres Bello.

{\it Facilities:} \facility{Very Large Telescope:Kueyen (FLAMES)}, \facility{HST (ACS)}, \facility{Magellan:Baade (IMACS)}.

\appendix
\section{Testing for Potential Data Analysis Biases}
\label{sec:test_databias}
We investigated the potential of erroneous literature measurements giving rise to the \hdyn\ sequence in Figure~\ref{fig:mass_mld_size}, and find that one object may be explained by discrepancies found in the literature.~For GC\,0225, we took the apparent magnitude $V\!=\!19.93$ mag from the \cite{woo07} catalogue, which is 2.86 magnitudes fainter than in the discovery publication of \cite{hol99}.~We do not attempt to explain this discrepancy, but note that if the brighter measurement is used, it leads to $\Upsilon_V^{\rm dyn}$ a factor of $\sim\!14$ lower, bringing it more in line with the \ldyn\ sequence.~Additionally, \cite{mie13} note an inconsistency in the size measured by \cite{hol99}, suggesting that $r_h$ is $\sim\!3.8\times$ smaller than originally estimated.~The smaller size would naturally lower $\Upsilon_V^{\rm dyn}$ by the same factor, bringing it more in line with the two intermediate (but still $\ga10$) $\Upsilon_V^{\rm dyn}$ clusters.~If GC\,0225 is both brighter and smaller, then its $\Upsilon_V^{\rm dyn}$ would approach one. Eliminating it from the power-law fit results in virtually the same slope ($\alpha\!=\!0.78\!\pm\!0.06$), with slightly larger scatter.

We also tested for various data reduction effects that may artificially give rise to the \ldyn\ and \hdyn\ sequences.~For example, a straight-forward explanation is that in performing the convolutions to estimate $\sigma_{\rm ppxf}$, the {\it ppxf} code may have ``jumped'' over the targeted Mg$b$ and Fe\,5270 spectral features and based the kinematics on the wrong combination of spectral lines due to the relatively low-S/N of some spectra.~This explanation would require an error on $v_{r,{\rm ppxf}}$ of $\ga\!100$ km\,s$^{-1}$ which Figure~\ref{fig:vrcomp} indicates is not present in our data.~Still, given that multiple $v_r$ measurements can be found in the literature for many of our sample GCs, we individually investigated the \hdyn\ sequence objects to search for any literature $v_r$ estimates that are discrepant by $\ga\!100$ km s$^{-1}$.~We found only three such clusters (GC\,0115, GC\,0225, and GC\,0437 by $\Delta v_{r,{\rm max}}=$142, 93, and 110 km s$^{-1}$, respectively), noting that they each fall within the error bars of at least one literature value. Nonetheless, even if all of three GCs are omitted from the fit shown in Figure~\ref{fig:mass_mld_size}, the relation again does not change significantly ($\alpha\!=\!0.78\!\pm\!0.06$).

\section{Testing for Target Confusion}
\label{sec:test_confusion}
The significant luminosity difference between the \ldyn\ and \hdyn\ sequence GCs calls for investigations into whether the bifurcation is due to observational effects.~In general, it is unlikely that any contamination of the sampled GC flux by other sources would produce two such relations, instead of just increased scatter. However, to address this issue we have visually inspected all available archival {\it HST} imaging data for any potential contamination of the \hdyn\ sequence GCs and found no indications for any target confusion due to foreground starlight, background galaxies, or enhanced surface brightness fluctuations of the surrounding galaxy light for each GC.

\section{Testing for Spurious Results Due to Noisy Spectra}
\label{sec:test_noise}

\begin{figure}
\centering
\includegraphics[width=8.9cm, bb=0 0 900 900]{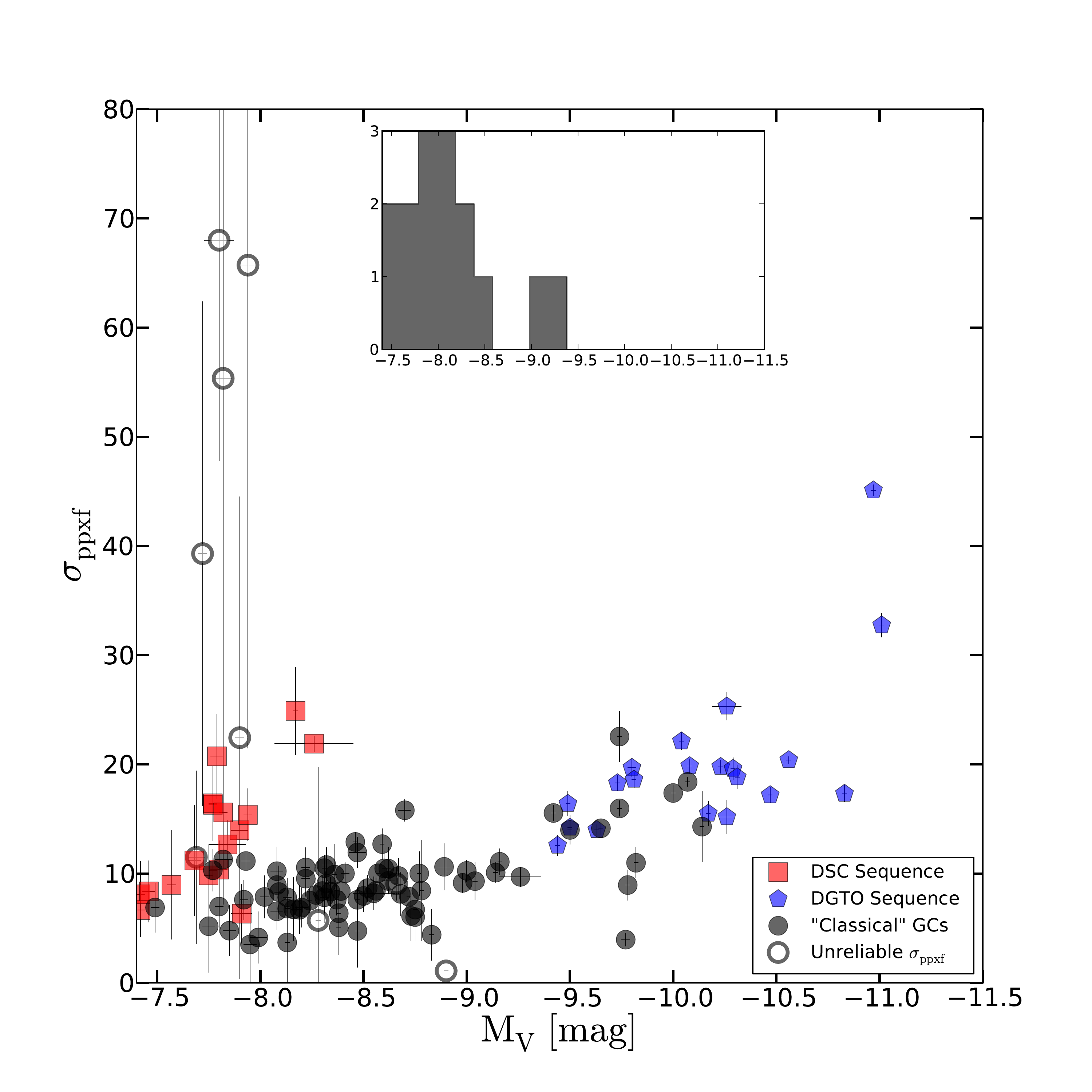}
\includegraphics[width=8.9cm, bb=0 0 760 760]{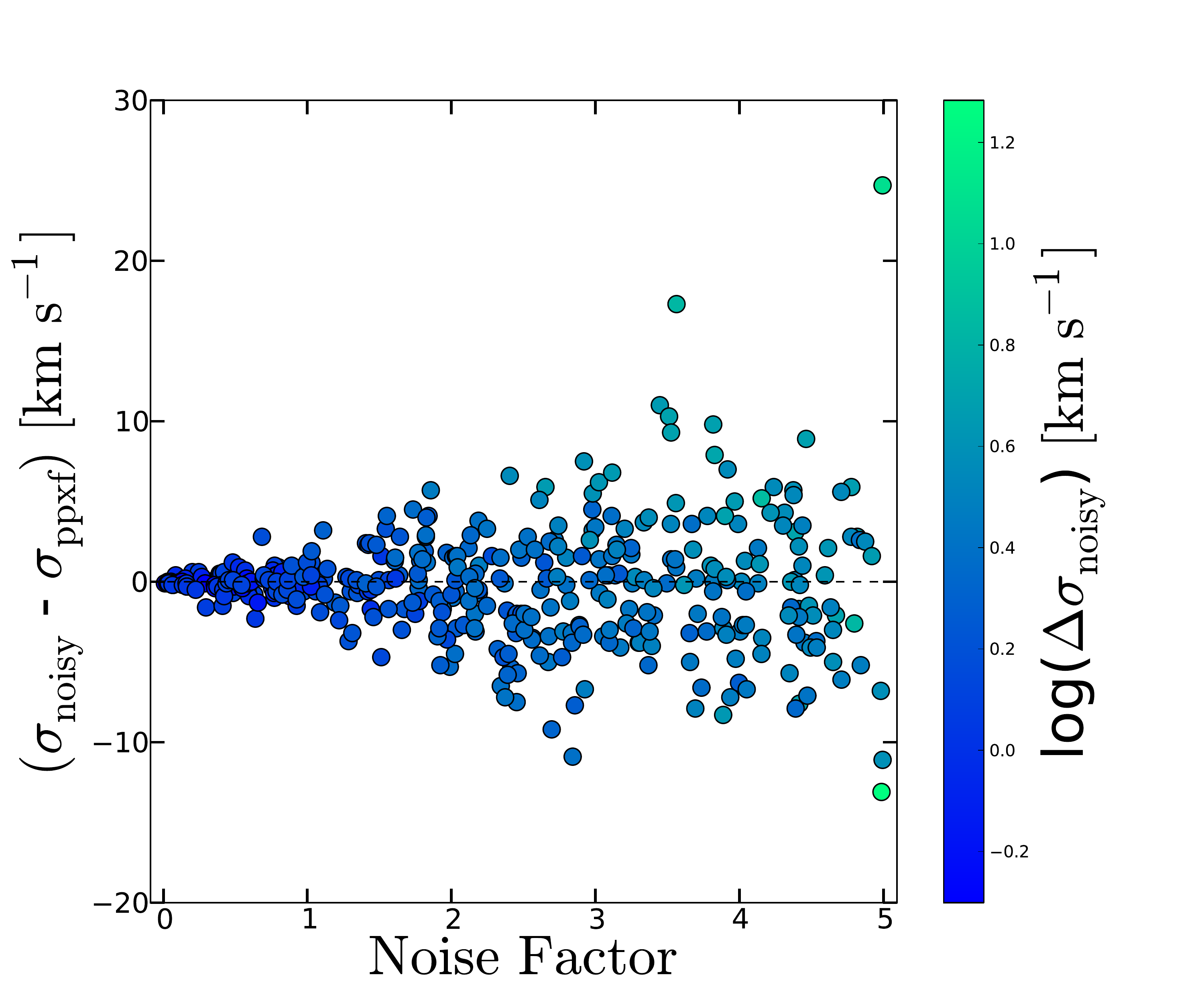}
\caption{{\it (Left panel):}~$\sigma_{\rm ppxf}$ values as a function of  $M_V$, with brightness increasing to the right.~Symbol shapes are as in Figure~\ref{fig:mass_mld_size}, with the addition of objects for which $\sigma_{\rm ppxf}$ could not be estimated (grey inset histogram) or which we considered unreliable (open grey circles).~{\it (Right panel):}~Difference between velocity dispersion measured on spectra with artificially added noise, $\sigma_{\rm noisy}$, and our measurements, $\sigma_{\rm ppxf}$, as a function of the amount of noise added in units of the standard deviation of the original spectra. The errors of the $\sigma_{\rm noisy}$ measurements are shown on a logarithmic scale, parametrized by the color shading to illustrate the increased uncertainties as more noise is added.~If poor-quality spectra were the cause of the \hdyn\ sequence, an upturn of the $\sigma_{\rm noisy}\!-\!\sigma_{\rm ppxf}$ difference would be expected towards larger noise factors; however, the symmetric increase in spread to both negative and positive $\sigma_{\rm noisy}\!-\!\sigma_{\rm ppxf}$ values clearly rules out this possibility.}
\label{fig:noise_test}
\end{figure}

The well-defined \hdyn\ relation could be due to the observational limit of our data. The solid black line shown in Figure~\ref{fig:mass_mld_size} shows where GCs with $M_V=-7.4$\, mag (the faintest measured objects in our sample) would lie.~Objects to the left are inaccessible to our survey, and the nearly parallel alignment of this line to the \hdyn\ sequence suggests that the small scatter may be an artifact of this limit.~Indeed, the left panel of Figure~\ref{fig:noise_test} shows $\sigma_{\rm ppxf}$ as a function of $M_V$ for our entire sample with symbols as in Figure~\ref{fig:mass_mld_size}, but GCs with unreliable $\sigma_{\rm ppxf}$ shown as open grey circles, and the inset histogram showing the $M_V$ distribution of GCs for which {\it ppxf} could not derive a $\sigma_{\rm ppxf}$ estimate.~It can be seen that most objects with poor/unavailable $\sigma_{\rm ppxf}$ encroach upon the luminosities of the \hdyn\ sequence.~These objects are possibly ``classical'' GCs with absorption features too narrow for GIRAFFE to resolve, and/or for {\it ppxf} to accurately measure through the noise, whereas \hdyn\ objects have features sufficiently broadened to be measurable.~This effect would give the false impression that almost all objects of $M_V\!\ga\!-8.0$\,mag seem to have anomalously high $\Upsilon_V^{\rm dyn}$.~In fact, the objects on the \hdyn\ sequence (red squares) show a trend of higher luminosity with larger $\sigma_{\rm ppxf}$, contrary to the expectation if noise were ``washing-out'' the finer spectral details used to estimate $\sigma_{\rm ppxf}$.

Since the \hdyn\ sequence objects are among the faintest objects in our sample (see Figure~\ref{fig:mag_mld_size}), an obvious point of concern is that their spectra are among the noisiest.~We therefore performed the following test to check against the potential of noise ``washing out'' some of the finer spectral details used to estimate $\sigma_{\rm ppxf}$, thus tricking the code into estimating systematically wider dispersions.~We took four \ldyn\ sequence GCs (GC\,0106, GC\,0277, GC\,0306 and GC\,0310; chosen to cover a reasonable luminosity range), artificially added varying levels of noise, and repeated the $\sigma_{\rm ppxf}$ measurements.~For each GC we added noise, pixel by pixel, by randomly drawing from a $N(0,n\cdot\sigma_{\rm spec})$ distribution where $\sigma_{\rm spec}$ is the flux dispersion intrinsic to each spectrum, and $n$ is drawn from $U(0,5)$ distributions.~This procedure decreases the spectral S/N by a factor $1/n$.~Repeating this process 100 times for each GC then builds a picture of how well behaved the {\it ppxf} code is for increasingly noisy spectra.~The right panel of Figure~\ref{fig:noise_test} shows the results of this exercise for all four GCs and it is clear that even for large amounts of noise, the points cluster well around the adopted $\sigma_{\rm ppxf}$ values, with increased dispersion bracketing the nominal values accompanied by larger errors.~If sudden ``jumps'' in $\sigma_{\rm ppxf}$ due to noisy spectra were the cause of the elevated $\Upsilon_V^{\rm dyn}$ shown by the \hdyn\ sequence, then it would be expected that the points in Figure \ref{fig:noise_test} would tend towards higher $\sigma_{\rm noisy}$ estimates with stronger amplified noise.~No hint of such asymmetric trend is seen in the plot, and we conclude that the \hdyn\ sequence is not artificially created by poor quality spectra.

\begin{figure}
\centering
\includegraphics[width=8.9cm, bb=0 0 864 864]{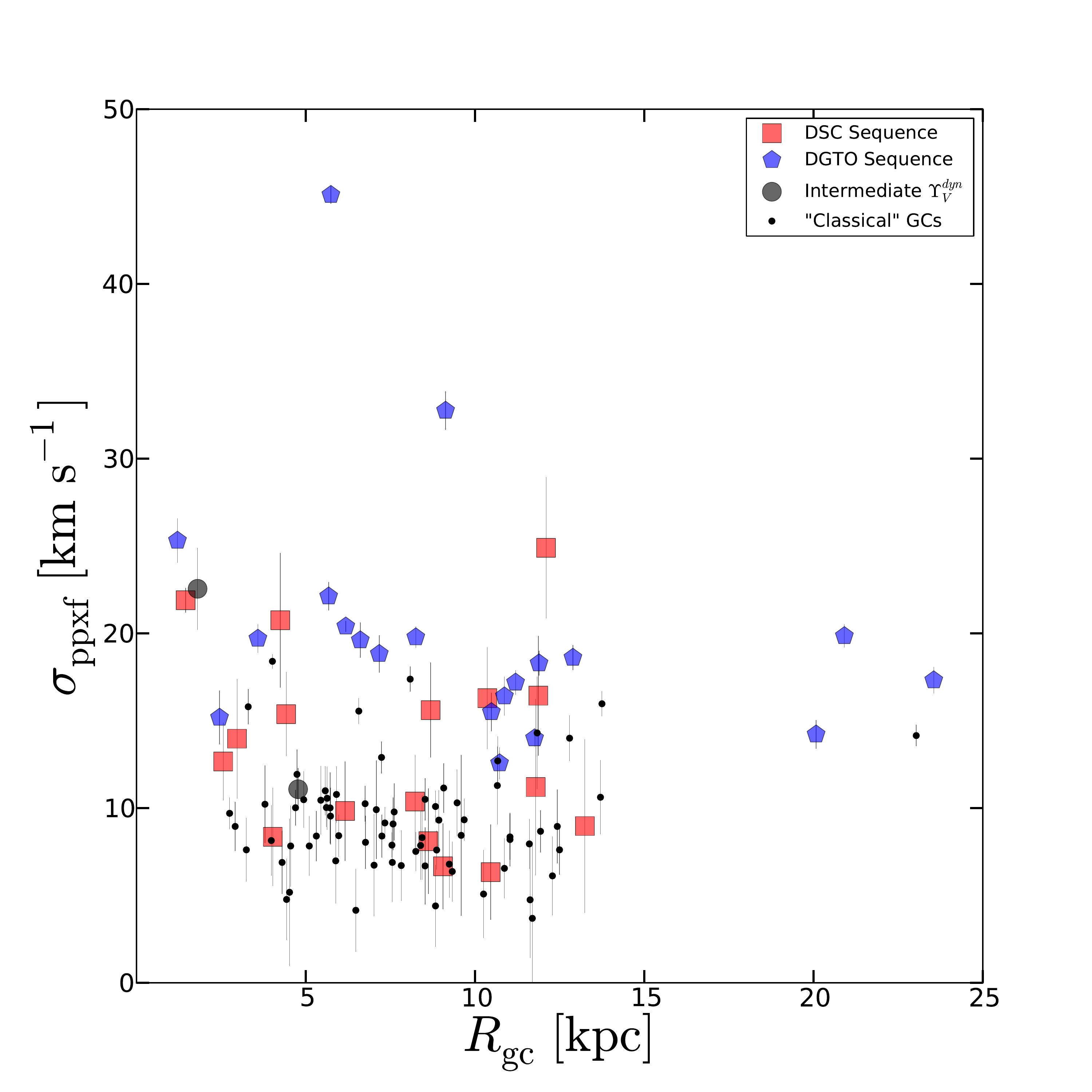}
\includegraphics[width=8.9cm, bb=0 0 864 864]{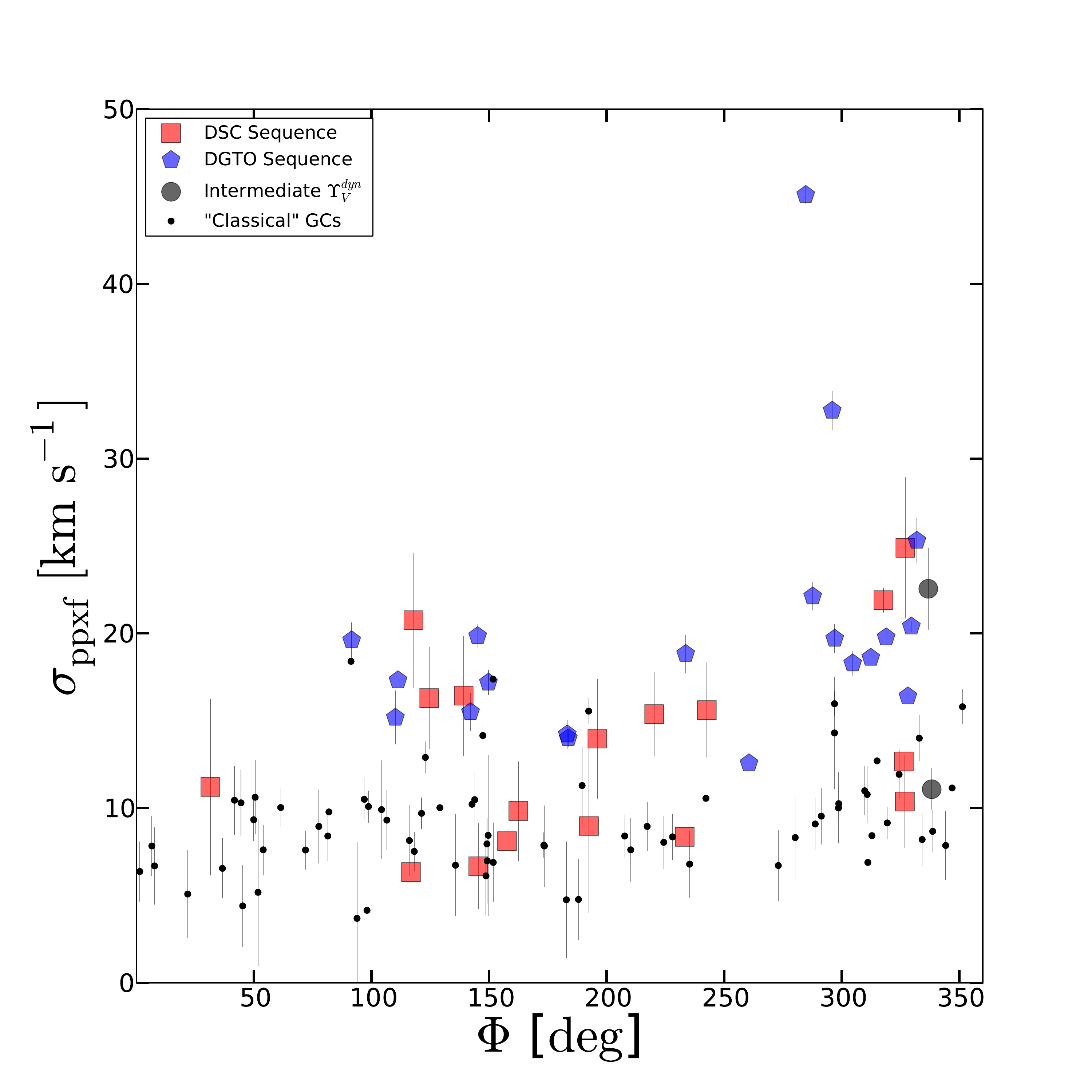}
\caption{{\it (Left panel)}:~Raw $\sigma_{\rm ppxf}$ measurement shown as a function of galactocentric radius, $R_{\rm gc}$, with symbols as in Figures \ref{fig:mass_mld_size} and \ref{fig:mag_mld_size}. {\it (Right panel)}:~Same as in the left panel, but showing $\sigma_{\rm ppxf}$ as a function of azimuthal angle, with 0$^\circ$ corresponding to North and increasing clockwise (see also Figure~\ref{fig:n5128_targs}). No obvious correlations between $\sigma_{\rm ppxf}$, $R_{\rm gc}$, or $\Phi$ are seen, effectively ruling out the possibility of systematic variations of $\sigma_{\rm ppxf}$ with spatial distribution giving rise to the \ldyn\ or \hdyn\ sequences.}
\label{fig:sigma_radial}
\end{figure} 

\section{Testing for Correlations with Galactocentric Radius and Azimuthal Angle}
\label{sec:test_galrad}
In searching for the origins of the structures found in Figure~\ref{fig:mass_mld_size}, in particular the bifurcation at high ${\cal{M}}_{\rm dyn}$, we test whether any correlation of the measured $\sigma_{\rm ppxf}$ with spatial distribution might give deeper insight into our results. For the following, we note that we are using the raw $\sigma_{\rm ppxf}$ measurement directly obtained from the spectra without subsequent correction for GC light profile sampling.~We plot in Figure~\ref{fig:sigma_radial} $\sigma_{\rm ppxf}$ as functions of galactocentric radius ($R_{\rm gc}$; left panel), and azimuthal angle ($\Phi$; right panel), where 0$^\circ$ corresponds to north and $\Phi$ increases clockwise.~We find no indications for any $\sigma_{\rm ppxf}$-$R_{\rm gc}$ or $\sigma_{\rm ppxf}$-$\Phi$ correlations for our GC sample. More importantly, we find no such correlations for either of the GC sub-samples, in particular for the \hdyn\ GCs (red squares).~We conclude that systematic variations in $\sigma_{\rm ppxf}$ as a function of spatial distribution are not responsible for the bifurcation in the $\Upsilon_V^{\rm dyn}$-${\cal{M}}_{\rm dyn}$ relation.

\begin{figure*}
\centering
\includegraphics[width=8.9cm, bb=0 0 930 864]{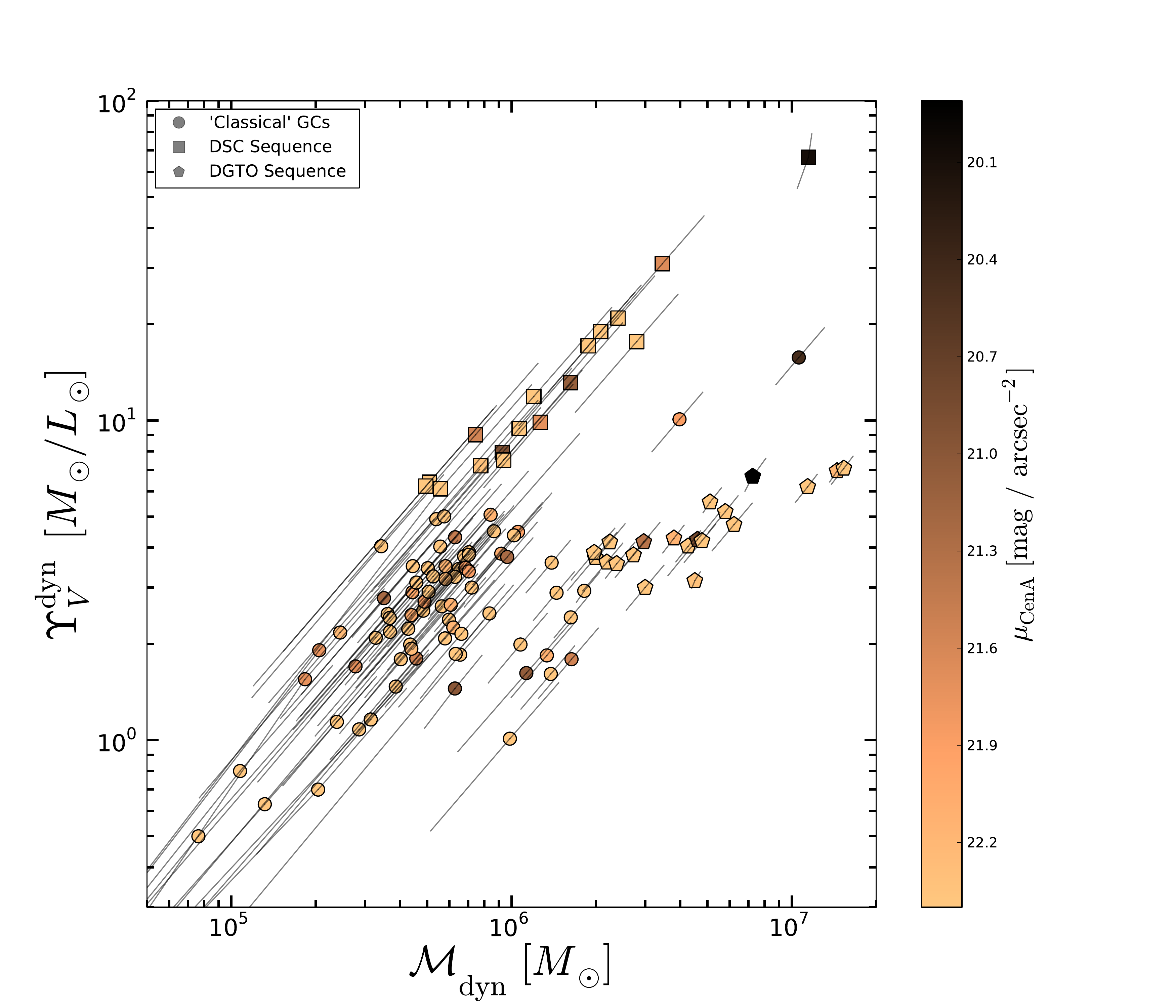}
\includegraphics[width=8.9cm, bb=0 0 930 864]{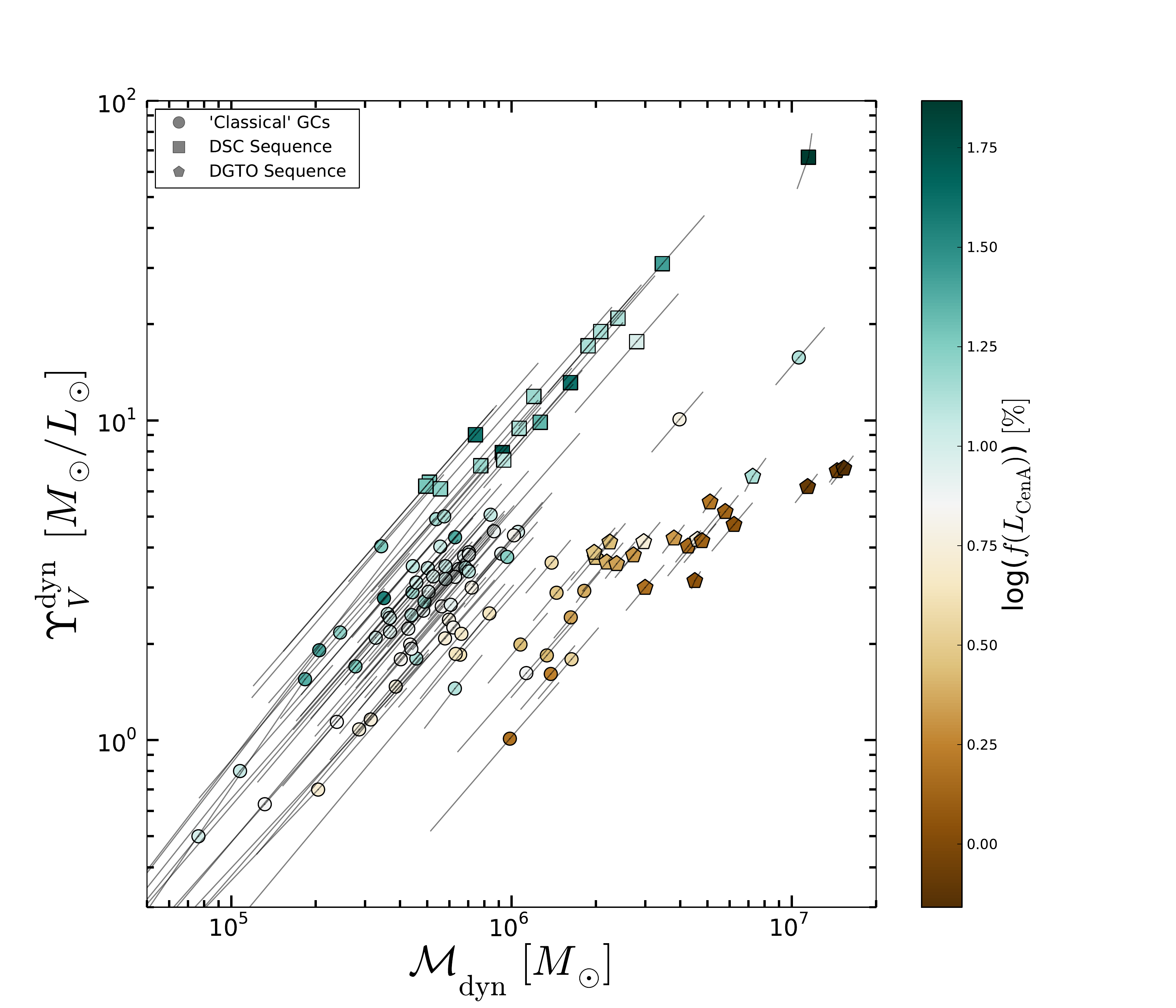}
\caption{$\Upsilon_V^{\rm dyn}$-${\cal{M}}_{\rm dyn}$ plane as in Figure~\ref{fig:mass_mld_size}, but now in the left panel  with the local \cena\ surface brightness parametrizing the data point color, and in the right panel with the decadal logarithm of the fractional \cena\ galaxy light contribution to each of the sampled GC fluxes, {\it before} sky subtraction.}
\label{fig:contamination}
\end{figure*} 

\section{Testing for Galaxy Surface Brightness Contamination}
\label{sec:test_sky}

Here we investigate the fractional flux from \cena's diffuse galaxy light component that enters each fibre together with the GC light.~We consider the potential impact of such fibre contamination, which could potentially lead the {\it ppxf} code to predict artificially high $\sigma_{\rm ppxf}$ values. However, if this was the case, one would expect that the associated error bars would reflect this uncertainty (given full error propagation), and/or that the elevated $\Upsilon_V^{\rm dyn}$ values required to explain the \hdyn\ sequence (see \S \ref{sec:origins}) would either tend towards {\it higher} values with {\it decreasing} luminosity, which is not seen in our data (see Figure~\ref{fig:mag_mld_size} and \ref{fig:xmass_frac}), or that there would be a vanishing or no $\Upsilon_V^{\rm dyn}$-${\cal{M}}_{\rm dyn}$ correlation at all. As can be seen in Figure~\ref{fig:mass_mld_size}, the \hdyn\ sequence is consistent with the individual error bars, which themselves are dominated by the fully propagated uncertainties on $\sigma$.

Given that our background light SED modeling and subsequent subtraction from each fibre is not a localized process (see \S~\ref{sec:skysub}), visual inspection of the sky+background subtracted GC spectra indicated that they should not be significantly affected by residual galaxy light contamination. To be sure, we explicitly calculated the expected residuals by comparing the difference between the flux of \cena\ at the location of each target and our final sky estimates to the flux of our final target spectra. The results suggested that residual components due to diffuse galaxy light is for all GCs $<\!4$\%, and in most cases near the 1\% level, with the expected trend toward lower residuals with increasing $R_{\rm gc}$. We then tested for the maximal effects of these residuals on the \hdyn\ sequence objects by adding the signal from the nearest sky fibre to the final, reduced target spectra at flux levels between 1\% and 90\% of the object flux and re-measuring $\sigma_{\rm ppxf}$. The results showed that below $\sim20\,\%$, the effect on $\sigma_{\rm ppxf}$ was negligible, above which $\sigma_{\rm ppxf}$ monotonically growing with increased sky contamination, as one would expect.

In any case, we plot in the left panel of Figure~\ref{fig:contamination} the $\Upsilon_V^{\rm dyn}$-${\cal{M}}_{\rm dyn}$ plane, as in Figure~\ref{fig:mass_mld_size}, but parametrize the data point color this time with the local surface brightness of \cena\ at each GC location.~We obtain the surface brightness values from the profiles measured as part of the Carnegie-Irvine Galaxy Survey \citep[CGS; see][]{ho11,li11}.~No obvious correlations with local surface brightness values are found that would drive the bifurcation in the $\Upsilon_V^{\rm dyn}$-${\cal{M}}_{\rm dyn}$ relation.~We continue by using the surface brightness profile for \cena\ to compute the fractional background flux, $f(L_{\rm CenA})$, entering each fibre with the GC fluxes.~The right panel of Figure~\ref{fig:contamination} shows the results, with color parameterized by $\log f(L_{\rm CenA})$ {\it before} any sky+background subtraction is performed.~Several results are shown: 1) The ``classical" and \ldyn\ sequence GCs reveal the expected correlation between GC mass (i.e.~luminosity) and $f(L_{\rm CenA})$ with a decreasing $f(L_{\rm CenA})$ sequence from $\sim\!30\%$ for the low-mass GCs to negligible fractions for the highest-mass clusters; 2) GCs on the \hdyn\ sequence show on average larger $f(L_{\rm CenA})$ than GCs on the \ldyn\ sequence as expected from the trends seen in Figure~\ref{fig:mag_mld_size}, which shows that \hdyn\ GCs have luminosities of $-8.5\la M_V\la-7.5$ mag; 3) The \hdyn\ sequence GCs show {\it no} correlation between $f(L_{\rm CenA})$ and $\Upsilon_V^{\rm dyn}$ or ${\cal{M}}_{\rm dyn}$, in contrast to the expectation if the $\Upsilon_V^{\rm dyn}$-${\cal{M}}_{\rm dyn}$ sequences were created by insufficient background light subtraction.

While Figure~\ref{fig:contamination} illustrates fibre contamination before any sky modeling and subtraction was carried out, we performed one final test to check for the worst effect that insufficient background subtraction could have.~We first identified higher-S/N GCs with $\sigma_{\rm ppxf}$ equivalent to that which would give a ``normal'' $\Upsilon_V^{\rm dyn}$ for a \hdyn\ sequence cluster. For example, given the \hdyn\ GC\,0324's $\sigma_{\rm ppxf,GC\,0324}\!=\!24.89\,{\rm km}\,{\rm s}^{-1}$ and $\Upsilon_{V,GC\,0324}^{\rm dyn}\!=\!14.07\,M_\odot\,L_\odot^{-1}$, we identified the \ldyn\ GC\,0050 with $\sigma_{\rm ppxf,GC\,0050}\!=\!10.09\,{\rm km}\,{\rm s}^{-1}$, which would give $\Upsilon_{V,GC\,0324}^{\rm dyn}\!=\!2.31\,M_\odot\,L_\odot^{-1}$.~Using three such combinations, we scaled the individual spectra of the bright GC, pre-sky subtraction, to the flux level of the fainter cluster, and added the nearest sky fibre's signal. Performing the sky subtraction as before (see \S~\ref{sec:skysub}) with the next three closest fibres then simulated the noise levels of the faint GC, but preserved the spectral details of the brighter. Measuring $\sigma_{\rm ppxf}$ on the degraded spectrum was, at worst, consistent with our adopted measurements, and even led to a slightly lower $\sigma_{\rm ppxf}$ for two of the experiments.~We conclude that improper background subtraction is thus very unlikely to inflate our $\sigma_{\rm ppxf}$ measurements of the \hdyn\ sequence objects.

\end{document}